\documentclass[]{tCPH2e}

\begin{document}
\doi{10.1080/0010751YYxxxxxxxx}
 \issn{1366-5812}
\issnp{0010-7514}

\jvol{00} \jnum{00} \jyear{2010} 

\markboth{Roee Ozeri}{Contemporary Physics}

\articletype{Tutorial}

\title{The trapped-ion qubit tool box}


\author{Roee Ozeri$^{a}$$^{\ast}$\thanks{$^\ast$Corresponding author. Email: ozeri@weizmann.ac.il\vspace{6pt}}
\\\vspace{6pt}
$^{a}${\em{Department of Physics of Complex Systems, The Weizmann Institute of Science, Rehovot, 76100, Israel}}}

\maketitle

\begin{abstract}
In this tutorial we review the basic building blocks of Quantum Information Processing with cold trapped atomic-ions. We mainly focus on methods to implement single-qubit rotations and two-qubit entangling gates, which form a universal set of quantum gates. Different ion qubit choices and their respective gate implementations are described.\bigskip

\bigskip
\bigskip

\end{abstract}

\section{Introduction}\label{Introduction}
Quantum information processing (QIP) requires the coherent manipulation of the joint quantum state of a register of $N$ qubits, in a $2^N$ dimensional Hilbert space. Some of the resulting states are highly entangled superpositions in which the superposition parts represent states that differ in many of the $N$ qubits. For useful computations $N$ will be mesoscopically or even macroscopically large. A superposition of two states that are macroscopically distinct was never experimentally observed and is, according to many people, even hard to envisage. The violent contrast between the superposition principle, governing the behavior of microscopic systems, and our daily life experience at macroscopic scales was already noted in the early days of quantum mechanics \cite{Schrodinger1935}. The effort of realizing a macroscopic quantum computer is thus also (and primarily) an effort to advance the superposition principle onto a macroscopic scale and hence it enjoys the unique position of having both practical as well as fundamental importance.

\begin{figure}[h] 
    {\includegraphics[width=0.5\textwidth]{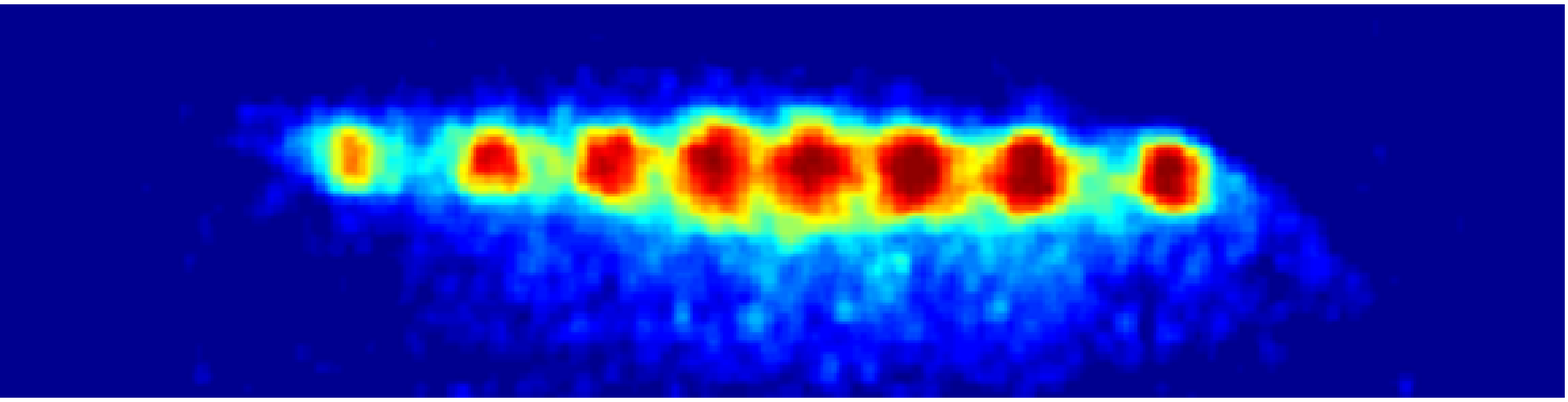}}
    \centering
    \caption{
    A linear crystal of eight trapped, and laser cooled $^{88}$Sr$^+$ ions.
    }\label{figure1}
\end{figure}

Among the many systems studied as a possible QIP platform, trapped-ions have many advantages. Atomic ions can be trapped by electric fields in ultra-high vacuum and then laser-cooled to extremely low temperatures. Figure \ref{figure1} shows a linear crystal of eight, trapped and laser-cooled, $^{88}$Sr$^+$ ions. Owing to their high degree of separation from any thermal environment, quantum superpositions of trapped-ions states have relatively long coherence times. Extremely accurate control has been demonstrated over their collective internal states \cite{Wineland_bible,LeibfriedRMP2003}. Trapped ions can be laser-cooled to the ground-state of their confining potential, thus providing excellent control of their motion as well. Qubits have been encoded into trapped-ion states \cite{Wineland_bible,LeibfriedRMP2003}. In the last decade or so, all the basic building blocks needed for coherent control of ion qubits were developed \cite{Home2009}. Several quantum algorithms have been demonstrated using up to eight or so ion qubits and, from a quantum computing point of view, the challenge that experimentalists in this field are currently facing is to find ways to scale this system up to large numbers.

The challenge of experimentally realizing a large-scale quantum computer is hard, and it has many different fronts. A large-scale trapping architecture has to be developed \cite{Kielpinski2002}, tools for applying parallel quantum gates in multiple trapping regions are needed \cite{Kim2009}, and so forth. One particular challenge is to be able to drive high-fidelity coherent operations on a collective quantum state of a large number of ion qubits. Obviously, the larger the quantum register is, the larger is the space of possible quantum operations. However, the number of possible operations in a quantum code has to be finite, and in particular, independent of the number of qubits in the register. This problem is analogous to the problem of efficiently compiling a computer program into a machine code that implements physical logical operations on a logical register of bits, in a classical computer. In QIP, a finite set of unitary operations, which can be concatenated to efficiently approximate any operation on an arbitrarily large qubit register, is called a universal quantum gate set. This tutorial reviews the basic building blocks of quantum computing operations with trapped-ion qubits, with an emphasis on ways to implement a universal quantum gate set.

\subsection{Universal gate set}\label{Universal gate set}
The perfect coherent control of a quantum state in a $2^N$ dimension
Hilbert space requires the ability, up to a global phase, to
unitarily connect any two state vectors in this space. This seems to be
a horrendous task, considering that any such unitary operator is
generated by $N^2-1$ independent generators. Fortunately, it was
shown that any such operation can be approximated to an arbitrary
precision by concatenating a finite number of operators chosen from a small (i.e. of size which is independent of $N$) set
\cite{Deutsch1989, Barenco1995}. This small and finite set of
operators is called a universal set of quantum gates.

\begin{figure}[h] 
    {\includegraphics[width=0.5\textwidth]{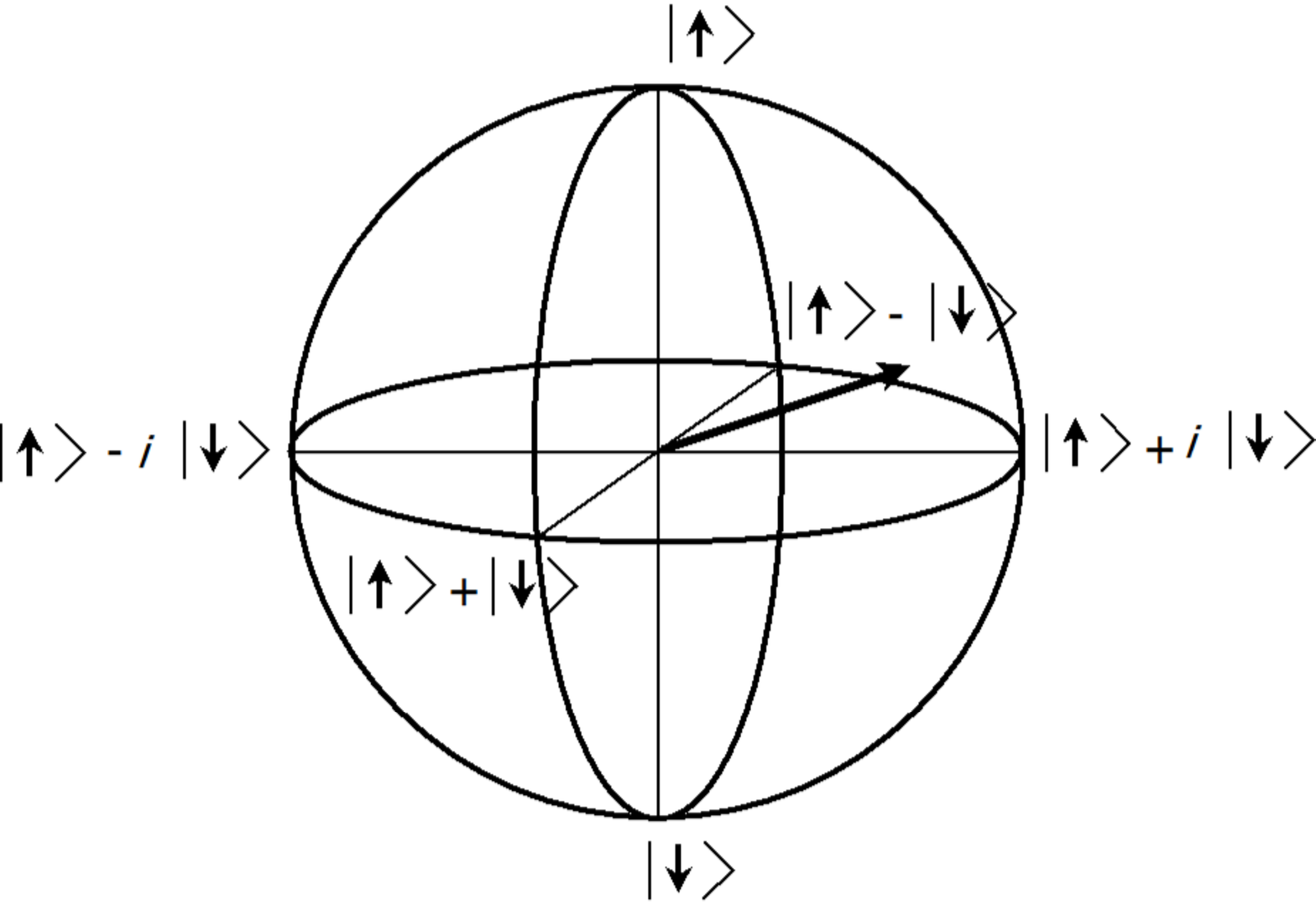}}
    \centering
    \caption{
    All pure states of a single qubit can be geometrically described as all the points on the surface of a sphere, called the "Bloch sphere". All possible single-qubit operations are therefore equivalent to rotations in three dimensions.
    }\label{figure2}
\end{figure}

Up to a global phase, a pure quantum state of a single qubit can be
visualized as a vector pointing at the surface of a sphere called
the Bloch sphere, illustrated in Fig.\ref{figure2}. Any unitary
single qubit operation is therefore equivalent to a rotation
$R\equiv R(\beta,\phi, \theta)$ of the Bloch vector. This rotation
is specified by the three Euler angles $\beta$, $\phi$, and $\theta$.
The angles $-\pi/2<\beta<\pi/2$, and $0<\phi<2\pi$ determine the
direction of $\bold{n}=(\cos{\beta}\cos{\phi},\cos{\beta}\sin{\phi},
\sin{\beta})$, the axis around which the Bloch vector is rotated, and $\theta$ is the rotation angle. In the
Hilbert space of a single qubit, rotations are represented by \emph{SU[2]}
operators, i.e. $2\times2$ unitary matrices with a unity determinant. These rotations are
generated by the three Pauli matrices $\hat\sigma_x$, $\hat\sigma_y$
and $\hat\sigma_z$. The rotation $R$ is therefore represented by the operator,
\begin{equation}
\hat{R}(\beta,\phi, \theta)=\hat{R}(\bold{n},\theta) =
e^{\frac{-i\bold{\hat{\sigma}}\cdot\bold{n}\theta}{2}} = \left [
   \begin{array}{cc}
   \cos\frac{\theta}{2}-in_z\sin\frac{\theta}{2}&(-in_x-n_y)\sin\frac{\theta}{2}\\
   (-in_x+n_y)\sin\frac{\theta}{2}&\cos\frac{\theta}{2}+in_z\sin\frac{\theta}{2}
   \end{array}\right ],
\label{Single qubit rotation}
\end{equation}
where $\bold{\hat{\sigma}}$ is the Pauli operators vector.

\begin{figure}[h] 
    {\includegraphics[width=0.5\textwidth]{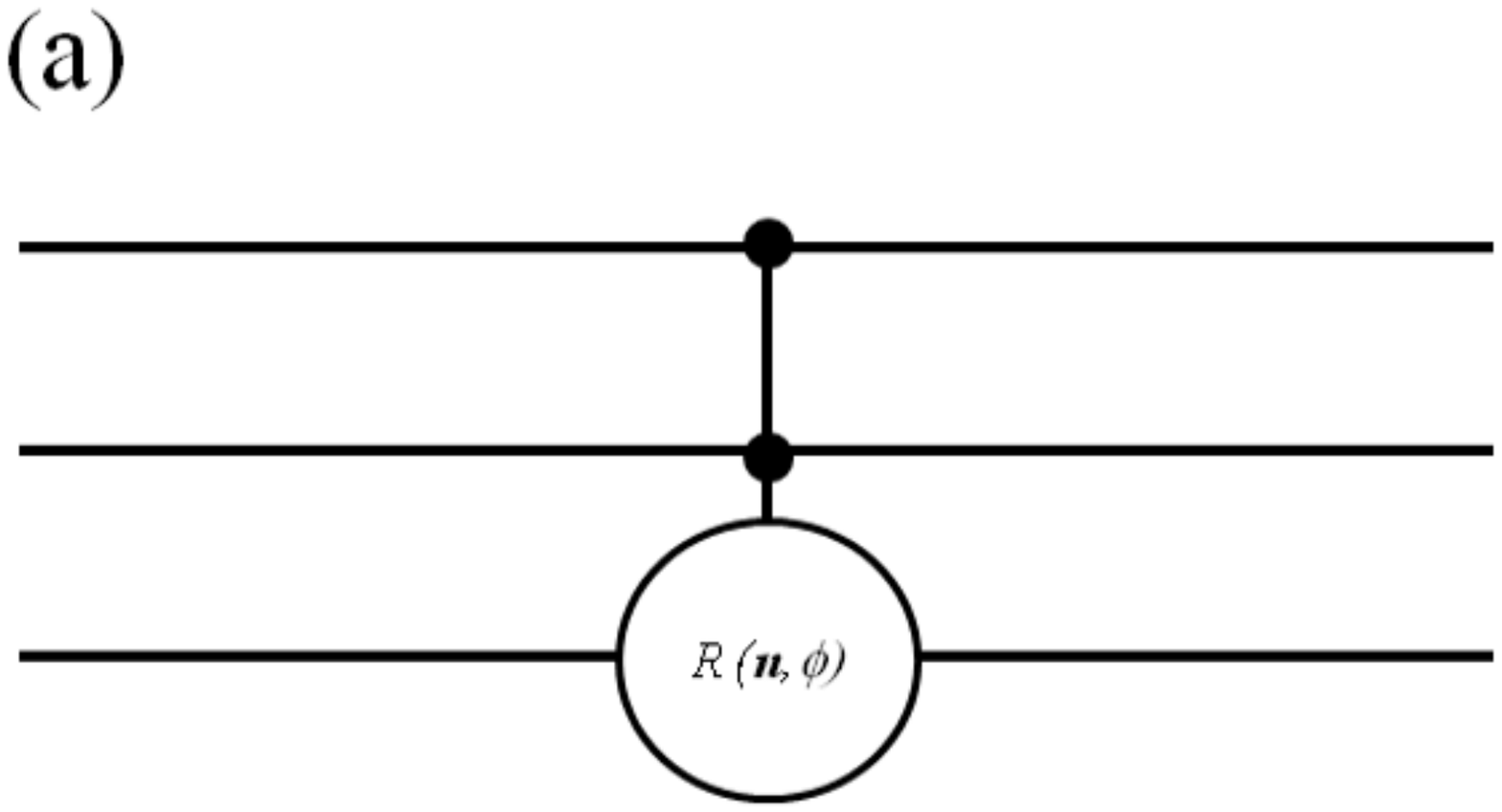}}
    {\includegraphics[width=0.5\textwidth]{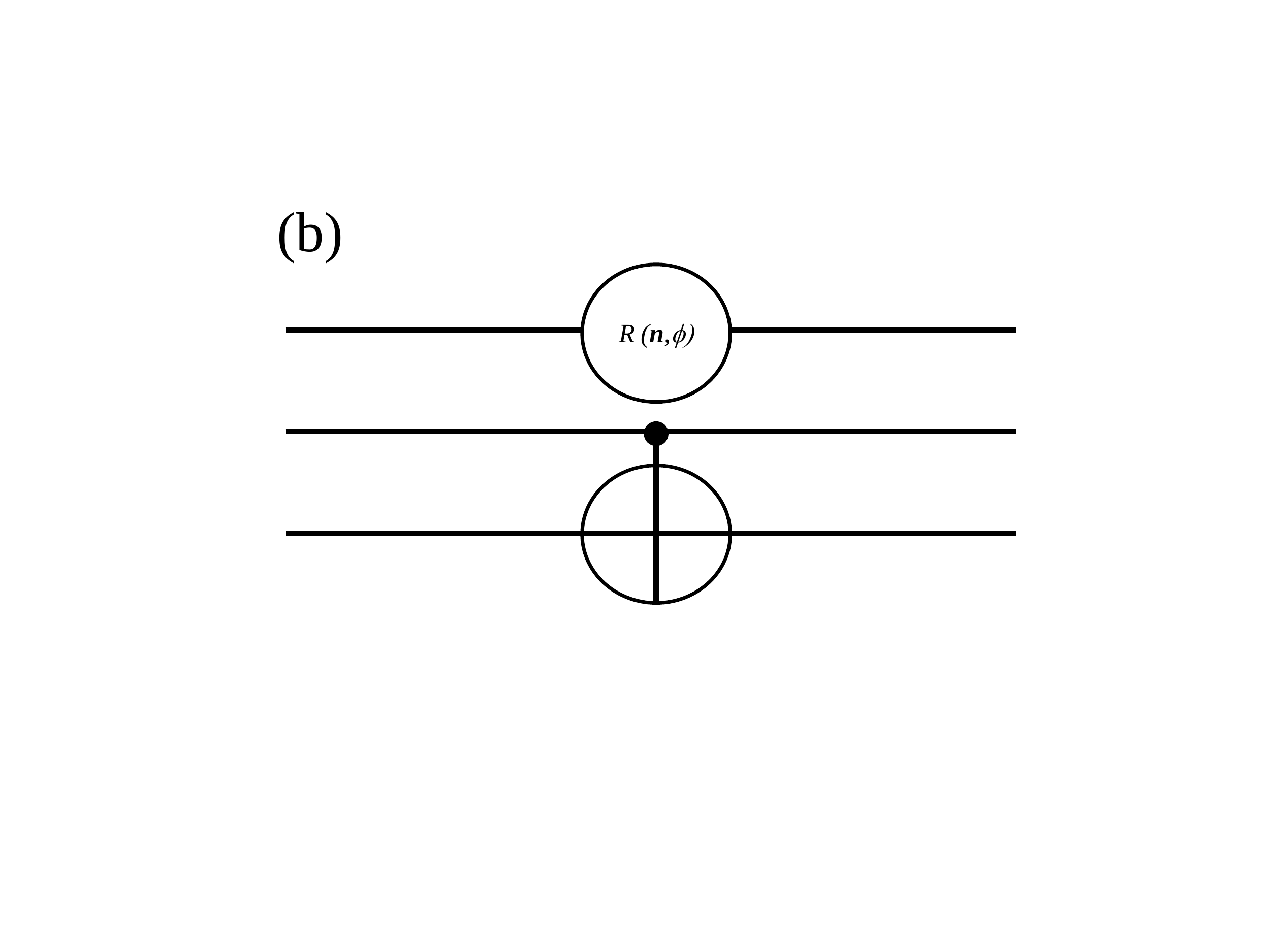}}
    \caption{
   Quantum circuits of universal gate sets. (a) The Toffoli gate is a controlled rotation. The state of a target qubit is rotated only if the state of both control qubits is in the logical $|\uparrow\rangle$. (b)Single-qubit rotations and controlled-not (CNOT). In a CNOT gate the state of a target qubit is rotated by $180^\circ$ around the $x$-axis, if the state of the control qubit is in the logical $|\uparrow\rangle$.
    }\label{figure3}
\end{figure}

Deutsch was the first to propose the quantum Toffoli gate
\cite{Deutsch1989}, a rotation $\hat{R}$ of a single target qubit
which is conditioned on the state of two control qubits, as a
universal gate set (see Fig. \ref{figure3}a). Here the state of the
target qubit undergoes a rotation $\hat{R}$ if and only if the state
of both control qubits is in the logical $|1\rangle$. Later, Barenco
and co-authors proved that single qubit rotations, $\hat{R}$,
together with two-qubit controlled-not (CNOT) gates, shown
diagrammatically in Fig. \ref{figure3}b, are a universal quantum
gate set as well \cite{Barenco1995}. A CNOT quantum gate performs a
$\theta=\pi$ rotation of the target qubit around the Bloch sphere
$x$ axis, i.e. $\hat{R}=\hat\sigma_x$, only if the control
qubit is in state $|1\rangle$. Both of these gate set examples are
not finite since $\hat{R}$ is defined through continuous parameters.
It has been shown, however, that by restricting to a small set of rotation
angles, any unitary operation in a $2^N$ dimension Hilbert space can
be approximated to accuracy $\epsilon$ by concatenating a string of
length $\log^c(N/\epsilon)$, where $c$ is a constant, approximately equal to 2, of operators from the universal gate sets
above \cite{Kitaev95,Ike&Mike}.

\subsection{QIP with trapped ions}\label{QIP with trapped ions}
Qubits that are encoded in the internal states of laser-cooled and
trapped ions are promising candidates as a quantum computing
platform \cite{Wineland_bible}. Trapped-ions are well isolated; thus
the coupling of environmental noise to their internal electronic
states is weak. The long coherence time attainable with atomic
superpositions has been long recognized and led to the development
of atomic time standards \cite{Essen1955}. Manipulation of
atomic superpositions with electro-magnetic radiation has been extensively studied \cite{Atom Photon interactions}. Single
atomic-qubit rotations were demonstrated long before the emergence
of quantum information science (e.g. \cite{Rabbi1947}).

In a seminal paper Cirac and Zoller proposed a method to implement a
CNOT gate between ion qubits by using their long-range Coulomb
interaction \cite{CiracZoller1995}. In the last few years, different
schemes for implementing quantum gates that can serve as parts of a
universal set were theoretically suggested \cite{SM1999,Milburn2000,Ripoll2003,Duan2006,Blatt2008} and experimentally
demonstrated on trapped-ion systems with high fidelity \cite{Liebfried2003,BenhelmNature2008}.

In this tutorial we review some of the current methods for implementing a
universal quantum gate set on trapped ion qubits. Here we will focus
on the universal gate set suggested in \cite{Barenco1995}, i.e.,
single qubit rotations and two-qubit CNOT gates. In Section \ref{The
ion qubit} we review different ion qubit choices, in Section
\ref{Single qubit gates} we outline methods of implementing single
qubit rotations on the various qubit choices. Finally, in Section
\ref{Two-qubit gates} we describe a method for implementing a two-qubit
entangling gate. Here we focus on a two-qubit phase gate that is
based on spin-dependent light forces.

\section{The ion qubit}\label{The ion qubit}
Quantum information is encoded in the internal electronic levels of
trapped ions. Ions with a single electron in the valence shell have
a relatively simple level structure and are therefore well suited
for this purpose. The relatively manageable electronic level
structure also simplifies the laser-cooling of such ``alkali-like''
ions. Once singly ionized, all the elements with stable isotopes in
group A II earth alkalies, Be$^+$, Mg$^+$, Ca$^+$, Sr$^+$, and
Ba$^+$, are good candidate ions with a single electron in the zero
angular momentum orbital, $S$, of the $n^{th}$ electronic level. All
of these ions have filled $S$ and $P$ orbitals at the $(n-1)$ level and
no electrons in the $D$ orbital of the $(n-1)$ level, if they are heavy enough
to have one. All the radioactively stable, singly ionized, group B
II transition metals, Zn$^+$, Cd$^+$ and Hg$^+$, qualify, with $(n-1)$
filled $S$, $P$, and $D$ orbitals and a single electron at the $n$
$S$ level. Finally, the only singly ionized Lanthanide with a single
electron in the valence shell is Yb$^+$, with filled $(n-1)$ $S$ and
$P$ and $(n-2)$ $F$ orbitals and a single electron at the $n$ $S$
level. The trapping and laser-cooling of all ions mentioned above
were demonstrated \cite{Wineland1978, Itano1985, Bollinger1985,
Schubert1989, Madej1990, Klein1990, Gudjons1995, Tanaka1997,
Matsubara2003}.
\begin{figure}[h] 
    {\includegraphics[width=0.5\textwidth]{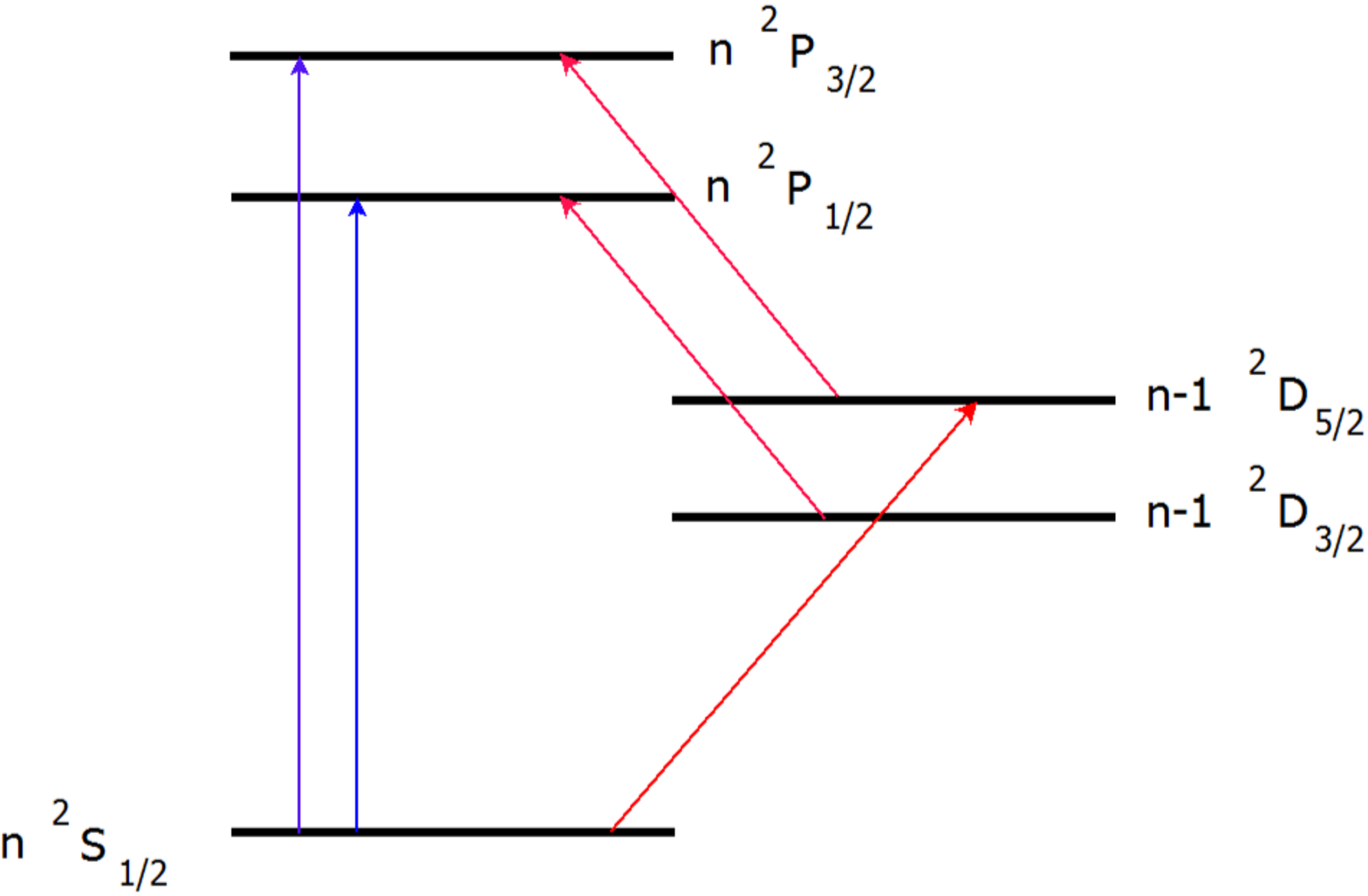}}
    \centering
    \caption{
  A diagram of the typical level structure of a singly ionized ``alkali-like'' ions. The ground electronic state is the zero angular momentum orbital, $S$, of the $n^{th}$ electronic level. An electric-dipole allowed transition to the $P$ orbital (typically in the blue-ultra-violet range) allows for laser-cooling, optical pumping and state-selective fluorescence detection. In those ions that are heavy enough to have a $D$ orbital in the $(n-1)^{th}$ electronic level, a narrow electric-quadrupole transition allows for the realization of an optical qubit.
    }\label{figure4}
\end{figure}

Since it is only singly ionized ``alkali-like'' ions that are
currently used for QIP purposes, it is worthwhile reviewing common
features in their internal level structure. Figure \ref{figure4} shows
a schematic diagram of the first few levels of such ``alkali-like''
ions. The ground electronic state of the valence electron is
$ns$ $^2S_{1/2}$, where $n$ is the valence shell principal quantum
number and $S$ is the zero angular momentum orbital. The electron
can be optically excited to the $P$ orbital via an electric-dipole
transition. Owing to spin-orbit coupling, the $P$ orbital splits into
two fine-structure levels: $np$ $^2P_{1/2}$ and $np$ $^2P_{3/2}$.
Table \ref{table1} lists the $S_{1/2}\rightarrow P_{1/2}$ and
$S_{1/2}\rightarrow P_{3/2}$ transition wavelengths and the natural
spectral width of the $P_{1/2}$ levels for the various ions.

Earth Alkaline ions heavy enough to have a $D$ orbital in the $(n-1)$
shell (Ca$^+$, Sr$^+$, and Ba$^+$, as well as Hg$^+$ and Yb$^+$),
have the $nd$ $^2D_{3/2}$ and $nd$ $^2D_{5/2}$ fine-structure levels
between the $S$ and the $P$ levels. Since the $S$ and the $D$ orbitals
both have even parity, the electric-dipole transition is forbidden
and the two levels are coupled only through their electric-quadrupole
moment. The lifetime of the $D$ levels (and hence the natural width
of the $S \rightarrow D$ transition) is therefore long (narrow), and
typically of the order of one second (one Hz). The  $D \rightarrow
P$ transition is an electric dipole transition and therefore, once excited to the $P$ level,
the electron can decay both to the states in the $S$ as well as in
the $D$ levels with a fixed branching ratio, $f$. The  $S
\rightarrow P$ transition is typically used for laser-cooling and
other QIP manipulations, and in those ions that have low-lying $D$
levels, re-pumping lasers, e.g. with a wavelength that is tuned to the $D \rightarrow P$
transition, are necessary to prevent the ion from
remaining in the long-lived $D$ manifold.

In odd isotopes or even isotopes with non-zero nuclear spin, $I\neq
0$; all the above levels are split into two or more hyperfine
manifolds. A small magnetic field usually splits these manifolds
further into different Zeeman sub-levels.

A qubit is best encoded in a pair of levels that are as resilient as
possible to decay and decoherence. Since levels in the $P$ manifold
have a very short lifetime (in the few $nS$ range), qubit levels are
typically encoded in two $S_{1/2}$ states or a $S$ and $D$
superposition. Three typical choices for an ion qubit are discussed
below.

\begin{table*}[tbp]
\begin{center}
  \begin{tabular}{|l|c|c|c|c|c|}
  \hline
  Ion&I&$\gamma/2\pi$ (MHz)&$\Delta_{hf}$ (GHz)&$\lambda_{1/2}$(nm)&$\lambda_{3/2}$(nm) \\
  \hline
  \hline
  $^{9}$Be$^+$& 3/2 & 19.6 & 1.25 & 313.1 & 313.0  \\
  \hline
  $^{25}$Mg$^+$& 5/2 & 41.3 & 1.79 & 280.3 & 279.6  \\
  \hline
  $^{43}$Ca$^+$& 7/2 & 22.5 & 3.23 & 396.8 & 393.4  \\
  \hline
  $^{67}$Zn$^+$& 5/2 & 62.2 & 7.2 & 206.2 & 202.5 \\
  \hline
  $^{87}$Sr$^+$& 9/2 & 21.5 & 5.00 & 421.6 & 407.8  \\
  \hline
  $^{111}$Cd$^+$& 1/2 & 50.5 & 14.53 & 226.5 & 214.4 \\
  \hline
  $^{137}$Ba$^+$& 3/2 & 20.1 & 8.04 & 493.4 & 455.4  \\
  \hline
  $^{171}$Yb$^+$& 1/2 & 19.7 & 12.64 & 369.4 & 328.9 \\
  \hline
  $^{199}$Hg$^+$& 1/2 & 54.7 & 40.51 & 194.2 & 165.0 \\
  \hline

  \end{tabular}\label{table1}
\caption{A list of atomic constants of several of the Hyperfine-ion
qubits. Here I is the nuclear spin, $\gamma$ is the natural width of
the $P_{1/2}$ level, $\Delta_{hf}$ is the $S_{1/2}$ hyperfine
splitting, $\lambda_{1/2}$ and $\lambda_{3/2}$ are the
$S_{1/2}\rightarrow P_{1/2}$ and $S_{1/2}\rightarrow P_{3/2}$
transition wavelengths respectively.}
\end{center}
\end{table*}

\subsection{Zeeman qubit}\label{Zeeman qubit}
Encoding a qubit in a pair of $S_{1/2}$ levels has the advantage of
a practically infinite spontaneous decay time. In ions where $I=0$,
the only states in the $S_{1/2}$ level are the two Zeeman states
corresponding to the valence electron spin pointing parallel and
anti-parallel to the external magnetic field direction
\cite{Lucas2003}. The energy splitting between the qubit levels is
$g_s \mu_B B \simeq 28$ MHz/mT. Here $g_s$ is the bound electron
$g$-factor, $\mu_B$ is Bohr magneton and $B$ is the magnetic field
magnitude. An advantage of choosing this qubit is that since the qubit
levels are the only levels in the $S_{1/2}$ manifold, optical
pumping out of the qubit manifold does not occur. One obvious
drawback of the Zeeman qubit is that since the energy difference
between the two qubit states depends linearly on the magnetic field,
magnetic field noise will cause dephasing. Furthermore, at low
magnetic fields the frequency separation between the two qubit
states is smaller than the $P$ level natural spectral width and
therefore state selective fluorescence cannot be directly applied
for qubit state detection.

\subsection{Hyperfine qubit}\label{Hyperfine qubit}
In odd isotopes or even isotopes with non-zero nuclear spin, $I\neq
0$, the $S_{1/2}$ is split into ($4I+2$) different $m_F$ states in two
hyperfine manifolds: $F = I\pm 1/2$. A qubit can be encoded in any
pair of hyperfine states \cite{Wineland1998, Blinov2004, Balzer2006,
Olmschenk2007, Benhelm2008}. The manifolds are split at a low
magnetic field by $\Delta_{hf}=(I+1/2)A_{hf}$, where $A_{hf}$ is the
ground-state hyperfine constant, which for all the ions discussed
here is in the few GHz range. Table \ref{table1} lists $\Delta_{hf}$ values
for various hyperfine ion qubits. A small magnetic field removes the
degeneracy between different Zeeman states in each of the two
manifolds. The Zeeman shift at a low magnetic field is given by
$g_F\mu_BBm_F$, and at all magnetic field levels it can be analytically
calculated from the Breit-Rabi formula \cite{Foot atomic physics}. A
clear advantage of the hyperfine qubit choice is that pairs of
levels can be found such that their energy separation does not
depend, to first order, on the magnetic field. The most commonly
used example is the $|F=I+1/2, m_f=0\rangle$ and $|F=I-1/2,
m_f=0\rangle$ states at zero magnetic field, often referred to as
the ``clock transition'' states. Working exactly at zero magnetic
field is not feasible since the degeneracy between different Zeeman
states has to be removed to enable spectroscopic addressing of
various transitions. Consequently, at a low non-zero field, $B_0$,
the magnetic field dependence of the clock transition is first order
but small. At certain magnetic fields, large enough such that the
magnetic dipole energy is comparable with $\hbar A_{hf}$, other
transitions can be found that are first order insensitive to
magnetic field variations \cite{Langer2006}.

\subsection{Optical qubit}\label{Optical qubit}
Ions that have a low-lying $D$ level have the possibility of
encoding a qubit into levels connected by the $S\rightarrow D$
optical transition. Since the lifetime of the $D$ level is typically
of the order of a second, the fundamental limit to the optical qubit
coherence time is long relative to a typical operation time.
However, in contrast with the two previously discussed qubit choices
where the local oscillator is in the Radio frequency (RF) or microwave range,
here the local oscillator is a laser. The coherence time of the
qubit is therefore also limited by a finite laser linewidth. A laser
linewidth comparable to the $D$ level natural spectral width ($\simeq 1$
Hz) requires laser fractional frequency stability of the order of
$10^{-14}$, a non-trivial task. This difficulty renders the optical
ion qubit choice less common. In ions with $I\neq 0$ both the $S$
and the $D$ levels are hyperfine splitted. Since the $P_{1/2}$ level
of ions cannot decay to the $D_{5/2}$, an optical qubit that is
encoded on the $S_{1/2}\rightarrow D_{5/2}$ transition enables
state-selective fluorescence on the $S_{1/2}\rightarrow P_{1/2}$
transition (see \ref{Initialization and detection}). Table
\ref{table2} lists the $S_{1/2}\rightarrow D_{5/2}$ transition
wavelengths, the $D_{5/2}$ level lifetime, and $f$, the branching
ratio of $P\rightarrow S / P\rightarrow D$ spontaneous decay for the
various ions.

\begin{table*}[tbp]
\begin{center}
  \begin{tabular}{|l|c|c|c|}
  \hline
  Ion&$\lambda_{D_{5/2}}$ (nm)&$D_{5/2}$ lifetime (sec)&$f$ \\
  \hline
  \hline
  Ca$^+$& 729.1 & 1.17 & 1/17 \\
  \hline
  Sr$^+$& 674.0 & 0.36 & 1/14\\
  \hline
  Ba$^+$& 1761.7 & 30 & 1/3 \\
  \hline
  Yb$^+$& 411.0 & 0.007 & 1/290 \\
  \hline
  Hg$^+$& 281.6 & 0.1 & 1/700 \\
  \hline

  \end{tabular}\label{table2}
\caption{The table lists the $S_{1/2}\rightarrow D_{5/2}$ transition
wavelengths, the $D_{5/2}$ level lifetime and, $f$, the branching
ratio of $P\rightarrow S / P\rightarrow D$ spontaneous decay in
various optical ion qubits}
\end{center}
\end{table*}

\subsection{Initialization and detection}\label{Initialization and detection}
All the ion qubit choices mentioned above are initialized to a
fiducial state using optical pumping techniques
\cite{Wineland_bible}. Qubit state detection is typically performed
using state-selective fluorescence \cite{Dehmelt1975, Itano1993,
Wineland_bible}. Photons are scattered from a laser beam resonant
with a transition connecting one of the qubit states to a short
lived state. The scattered light is collected on a photo-detector.
Whether or not the photo-detector measures light indicates the state
onto which the ion qubit superposition collapses. For optical ion
qubits, state-selective fluorescence is straightforward since the
two qubit levels are optically separated. For example, the ion
scatters photons from a laser beam that is resonant with the
$S_{1/2}-P_{1/2}$ transition \cite{Bergquist1987, SchmidtKahler2003,
Myerson2008}. The ion cannot decay from the $P_{1/2}$ to the
$D_{5/2}$ level and therefore if the qubit superposition collapses
on the $D_{5/2}$ state, then no photons are detected during the
detection period. Errors in detection here will be due to background
photon counts and the finite spontaneous decay probability of the
$D_{5/2}$ state during the detection period. Hyperfine qubit levels
are typically separated by much more than the $P$ level's natural
width (typically a few GHz as compared with a few tens of MHz; see
Table \ref{table1}). State-selective fluorescence is therefore
possible on a cycling transition, i.e., a transition in which, due to
selection rules, the $P$ level decays only to a single $S$ state
\cite{Itano1993, Acton2006}. Errors in this case will be mostly due to
off-resonance coupling of the detection laser to unwanted states
\cite{Langer2006}. A fairly large magnetic field is needed in order
to separate the two states of a Zeeman qubit by much more than the
$P$ level natural width. Hence, state-selective fluorescence cannot
be used directly. Detection is typically performed in Zeeman qubits
by shelving one of the qubit states on a spectrally distant
meta-stable state, e.g., one of the $D_{5/2}$ states
\cite{McDonnell2004, Wunderlich2007, Anna2011}. Other detection schemes that
use quantum logic protocols to enhance detection efficiency have been demonstrated as well \cite{Schaetz2005, Hume2007}.

\section{Single qubit gates}\label{Single qubit gates}
Single-ion qubit gates have been extensively reviewed (e.g. \cite{Wineland_bible, LeibfriedRMP2003, Wineland
LesHouches2003}). Here we initially provide a general description of
single-ion qubit gates followed by a discussion of specific
implementations for the various qubit choices.

The qubit levels are coupled by electro-magnetic (e.m.) traveling
plane waves, i.e., we assume that the ion is placed in the far-field
of the e.m. radiation. This assumption does not always hold, however, and
there have been several proposals to implement ion qubit gates using
near-field e.m. fields \cite{Wineland_bible, Wunderlich2001, Chiaverini2008,
Ospelkaus2008, Wunderlich2009, Ospelkaus2011}. The qubit evolution is determined by the time
dependent Hamiltonian,
\begin{equation}
\hat{H}(t) = \hat{H}_0 + \hat{V}(t). \label{time dependent
Hamiltonian}
\end{equation}
Here $\hat{H}_0$ is the free Hamiltonian of a spin connected to a 1D
harmonic oscillator (h.o.),
\begin{equation}
\hat{H}_0 = \frac{1}{2}\hbar\omega_0\hat{\sigma}_z +
\hbar\omega_m(\hat{a}^\dagger\hat{a} + \frac{1}{2}),
\end{equation}
where $\omega_0$ is the qubit levels frequency separation and
$\omega_m$ is the h.o. frequency. The time-dependent periodic
potential $\hat{V}(t)$ is induced by coupling to the e.m. wave,
\begin{equation}
\hat{V}(t) = \hbar \Omega_0 \big{(} \hat{\sigma}^+ + \hat{\sigma}^-
\big{)} \cos(\bold{k}\hat{x} - \omega t + \phi).\label{time dependent
potential}
\end{equation}
The coupling constant $\Omega_0$ is also known as the Rabi frequency
and $\hat{\sigma}^{+/-}$ are the spin raising/lowering operators,
\begin{eqnarray}
\hat{\sigma}_- = \hat{\sigma}_x + i\hat{\sigma}_y =
|\downarrow\rangle\langle\uparrow|\\ \nonumber \hat{\sigma}_+ =
\hat{\sigma}_x - i\hat{\sigma}_y =
|\uparrow\rangle\langle\downarrow|. \label{RaisingLowering}
\end{eqnarray}
The wave-vector, frequency, and phase of the e.m. plane wave are
$\bold{k}$, $\omega$, and $\phi$, respectively.  The position operator
can be written in terms of the h.o. creation and annihilation
operators, $\hat a$ and $\hat a^\dagger$,
\begin{equation}
\bold{k}\hat{x} = kx_{eq} + kx_0(\hat{a}^\dagger + \hat{a}) \equiv kx_{eq} +
\eta (\hat{a}^\dagger + \hat{a}),
\end{equation}
The phase owing to the equilibrium position of the ion, $kx_{eq}$, can
be absorbed in $\phi$. Here $k$ is the projection of $\bold{k}$ along the trap direction, $x_0 = \sqrt{\hbar/2m\omega_m}$ is the h.o. ground-state width and $\eta=kx_0$ is the Lamb-Dicke parameter.
Moving to the interaction representation and using the rotating-wave
approximation (RWA), the interaction Hamiltonian is,
\begin{equation}
H_{int}(t) = \hbar\Omega_0/2\hat{\sigma}_+ \exp\left [
i\eta(\hat{a}e^{-i\omega_mt} + \hat{a}^\dagger e^{i\omega_mt})\right
] e^{i(\phi-\delta t)} + H.C..\label{single qubit gate interaction
Hamiltonian}
\end{equation}
The RWA keeps only terms that vary slowly in time and
contribute the most to time evolution under $\hat{H}_{int}$.
Expanding the exponent in Eq. (\ref{single qubit gate interaction
Hamiltonian}) we obtain terms oscillating at multiples of $\omega_m$.
When $\delta=s\omega_m$, with integer $s$, only one of these terms
is in resonance and dominantly contributes to the ion qubit time
evolution. Here the Rabi frequency is modified to,
\begin{equation}
\Omega_{n,n+s} = \Omega_{n+s,n} = \Omega_0|\langle
n+s|e^{i\eta(\hat{a}+\hat{a}\dagger)}|n\rangle| \equiv \Omega_0
D_{n+s,n},\label{single qubit generalized Rabi}
\end{equation}
indicating the coupling of the two qubit levels and the $n$ and
$n+s$ h.o. levels. The coupling is modified by, $D_{n+s,n}$, the
Debye-Waller factor. For the 1D harmonic oscillator the Debye-Waller
factor is analytically calculated to be,
\begin{equation}
D_{n+s,n} =
e^{-\eta^2/2}\eta^{|s|}\sqrt{\frac{n_<!}{n_>!}}L_{n_<}^{|s|}(\eta^2).\label{Debye-Waller}
\end{equation}

Three different cases are commonly used. The carrier transition occurs when $s=0$ and the two qubit levels are coupled without
changing the h.o. motional state. The Hamiltonian in the interaction
representation is given by,
\begin{equation}
\hat{H}_{carrier} =
\frac{\hbar\Omega_{n,n}}{2}(\hat{\sigma}_+e^{i\phi} +
\hat{\sigma}_-e^{-i\phi})\label{Hcarrier}
\end{equation}
The red sideband (RSB) transition is the transition for which
$\delta=-\omega_m$ ($s=-1$). Here, when the ion qubit spin
state is raised, a single h.o. quantum of motion is annihilated. The
interaction Hamiltonian for the RSB is,
\begin{equation}
\hat{H}_{RSB} =
\frac{\hbar\Omega_{n-1,n}}{2}(\hat{a}\hat{\sigma}_+e^{i\phi}
+\hat{a}^\dagger\hat{\sigma}_-e^{-i\phi}).\label{HRSB}
\end{equation}
This Hamiltonian is identical to the well-known Jaynes-Cummings
Hamiltonian in quantum-optics. The third case is that of the
blue-sideband (BSB) transition, where $\delta=+\omega_m$ ($s=+1$)
and the raising of the qubit spin is accompanied by creating a single h.o. quantum,
\begin{equation}
\hat{H}_{BSB} = \hat{H}_{int} =
\frac{\hbar\Omega_{n+1,n}}{2}(\hat{a}^\dagger\hat{\sigma}_+e^{i\phi}
+\hat{a}\hat{\sigma}_-e^{-i\phi} ).\label{HBSB}
\end{equation}
Note that both the BSB and the RSB transitions entangle the qubit
spin with its motional state.

When the deviation of the ion from its average position is much
smaller than the radiation wavelength,
$\eta\sqrt{\langle(\hat{a}^\dagger + \hat{a})^2 \rangle }\ll 1$, the
Debye-Waller factor (\ref{Debye-Waller}) can be expanded using
$\eta$ as a small parameter. In this regime, known as the Lamb-Dicke
regime, the Rabi frequencies of the three cases discussed above can be
approximated by,
\begin{eqnarray}
&&\Omega_{n,n} \simeq \Omega_0\left [1-(n+1/2)\eta^2 \right ] \\
\label{oCarrier}&&
\Omega_{n-1,n} \simeq \Omega_0\sqrt{n}\eta\\
\label{oRSB}&&\Omega_{n+1,n} \simeq \Omega_0\sqrt{n+1}\eta. \label{oBSB}
 \label{Appendix6}
\end{eqnarray}
The carrier Rabi frequency $\Omega_{n,n}$ is corrected by
$(n+1/2)\eta^2$, a correction that is small in the Lamb-Dicke
regime. The RSB and BSB Rabi frequencies depend on the ion motion to
a greater extent and are proportional to $\eta\sqrt{n}$ and
$\eta\sqrt{n+1}$ respectively. Note that the RSB Rabi frequency
vanishes for $n=0$ since no h.o. quanta can be further extracted.

\begin{figure}[h] 
    {\includegraphics[angle=-90, width=0.5\textwidth]{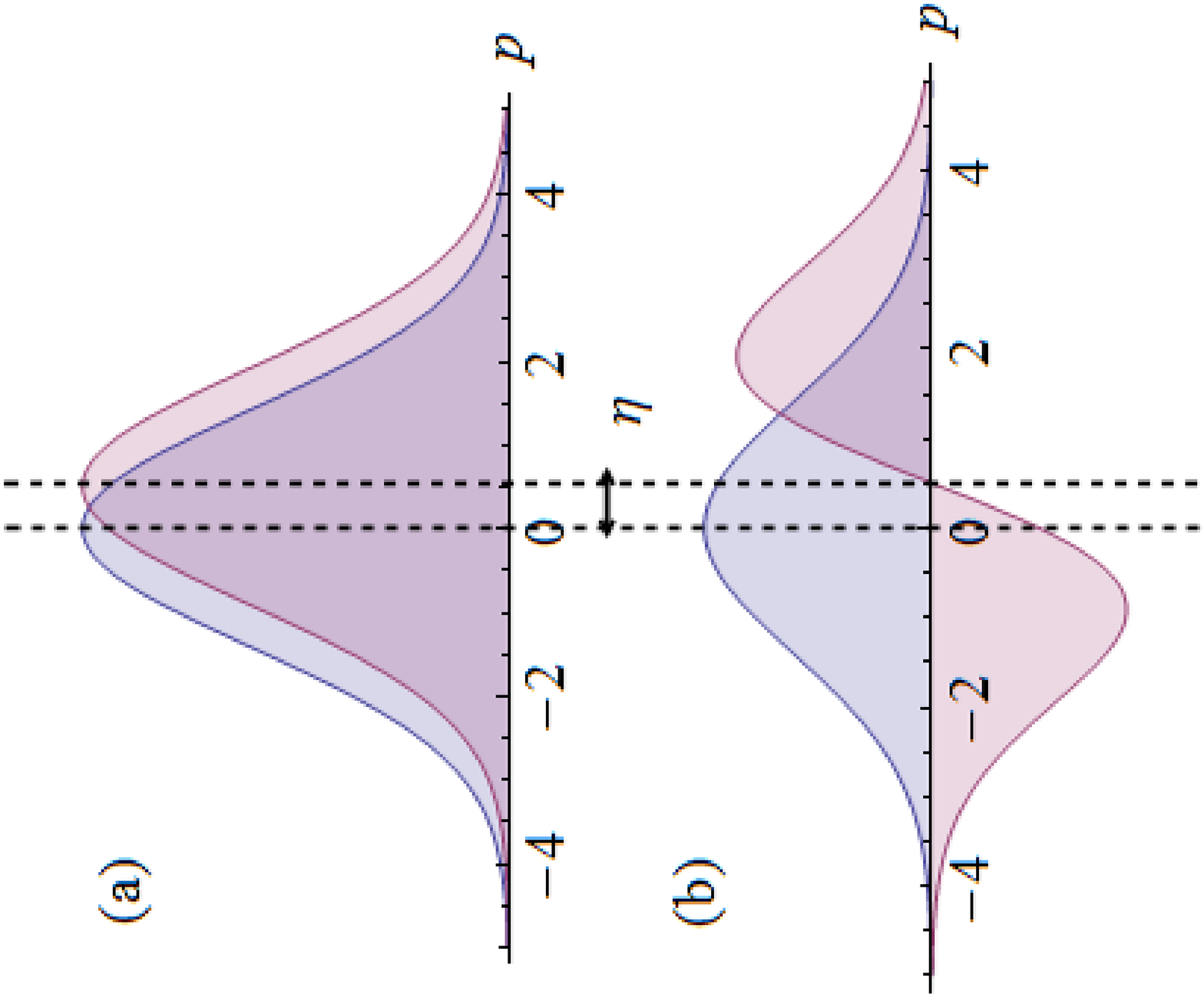}}
    \centering
    \caption{
 Momentum transfer in the Lamb-Dicke regime. (a) the overlap between a harmonic-oscillator (h.o.) ground-state and the same ground-state displaced by $\eta$, owing to photon recoil, decreases by $\eta^2/2$. Here, we use the units where $x_0$, the ground-state width equals unity. (b) The overlap between the h.o. first excited state and the recoil displaced ground-state increases linearly with $\eta$.
    }\label{figure5}
\end{figure}

The dependence of the Rabi frequencies on $\eta$ and $n$ can be
understood by considering the effect of photon recoil on the overlap between different h.o. wavefunctions. Looking at the overlap between a momentum-displaced
h.o. ground-state, a Gaussian, and either a ground-state (carrier
transition) or the first h.o. excited state (BSB transition), shown
in Fig \ref{figure5}. Here the wavefunction is plotted vs. momentum in
units of $2\pi\hbar/x_0$. In these units the h.o. ground-state width
is unity whereas the momentum displacement is $\eta \ll 1$. The overlap
between the ground-state and the displaced ground-state is
$1-\eta^2/2$ since the Gaussians are displaced by $\eta$ around
their maximal overlap point. The overlap between the displaced
Gaussian and the first excited state increases linearly with $\eta$
since the Gaussian peak is displaced along the excited state's linear
slope. The dependence on $n$ results from the fact that the distance
between nodes in the h.o. wavefunction, and therefore also the width
of each node is roughly $\hbar\sqrt{n}/x_0$. The overlap between
$n$ and the displaced $n+1$ wavefunctions would therefore also be
proportional to $\sqrt{n}$. The $\sqrt{n}$ dependence can also be
thought of as resulting from bosonic amplification.

To avoid entanglement of the ion qubit spin with its motion,
on-resonance carrier transitions are used to drive single qubit
gates. Here the interaction Hamiltonian, Eq.
(\ref{Hcarrier}), is time independent and the qubit time evolution
in the interaction representation is given by,
\begin{equation}
|\Psi(t) \rangle_{int} =
e^{-i\hat{H}_{int}t/\hbar}|\Psi(0)\rangle_{int} =
e^{-i\theta\vec{\sigma}\cdot\vec{n}}|\Psi(0)\rangle_{int}\equiv\hat{R}(\theta,\phi,0)|\Psi(0)\rangle_{int}.\label{carrier
propagator}
\end{equation}
Here $\theta=\Omega_{n,n}t$ and the Rabi vector direction are given
by the unit vector $\vec{n} = \left
(\begin{array}{ccc}\cos\phi,&\sin\phi,&0\end{array}\right )$
pointing in the Bloch sphere equatorial plane. The rotation matrix
$\hat{R}(0,\phi,\theta)$ represents the spinor rotation operator,
\begin{equation}
   \hat{R}(0,\phi,\theta)= \left [
   \begin{array}{cc}
   \cos\frac{\theta}{2}&-ie^{i\phi}\sin\frac{\theta}{2}\\
   -ie^{-i\phi}\sin\frac{\theta}{2}&\cos\frac{\theta}{2}
   \end{array}\right ] \label{rotation matrix}
\end{equation}

The origin of the coupling, represented by $\Omega_0$, can vary and
depends on the ion qubit choice. Next, we describe three
different single-qubit gate coupling implementations for the
different ion qubit types.

\subsection{Magnetic dipole coupling}\label{Magnetic dipole coupling}
\begin{figure}[h] 
     {\includegraphics[width=0.5\textwidth]{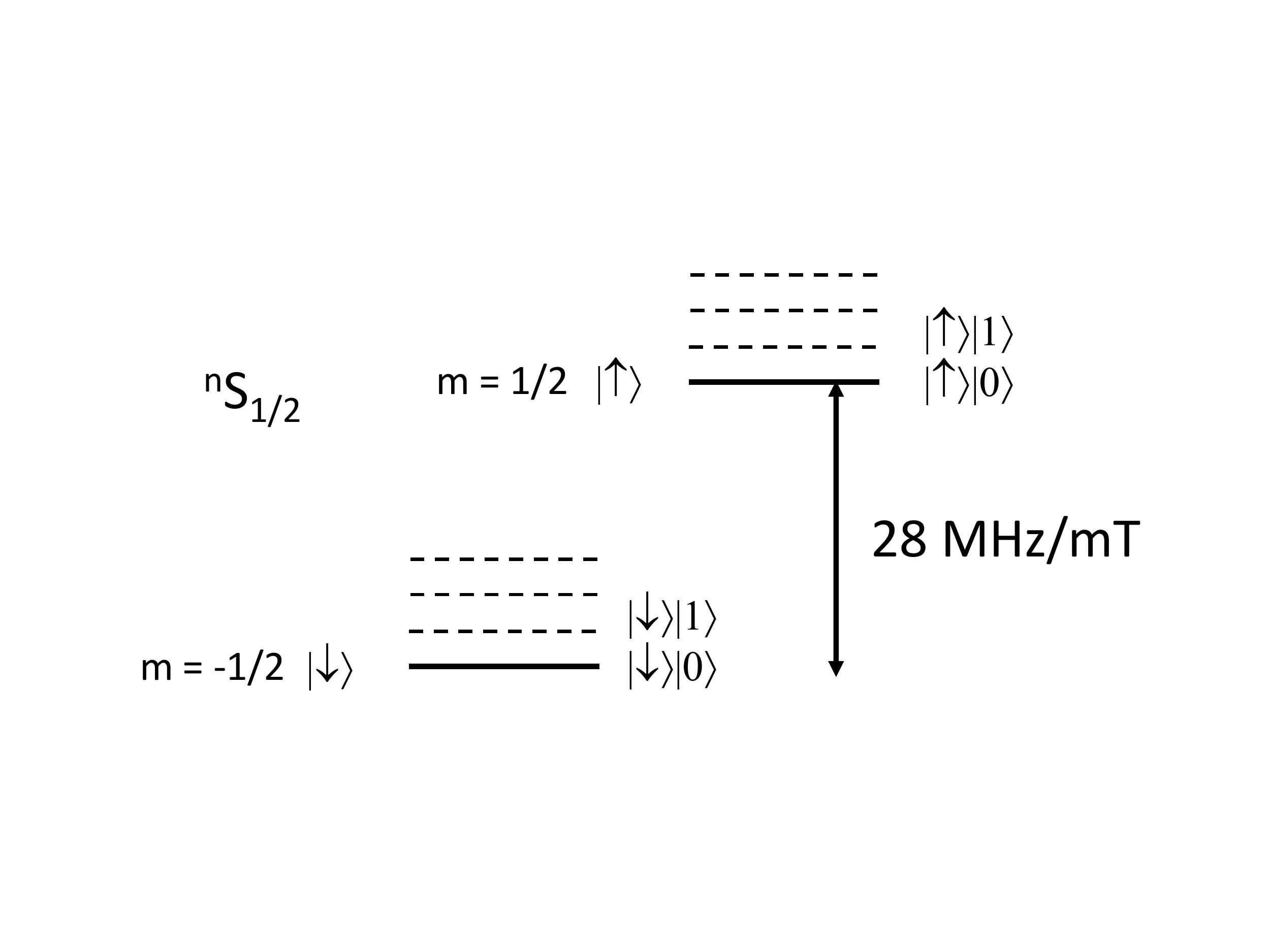}}
     {\includegraphics[width=0.5\textwidth]{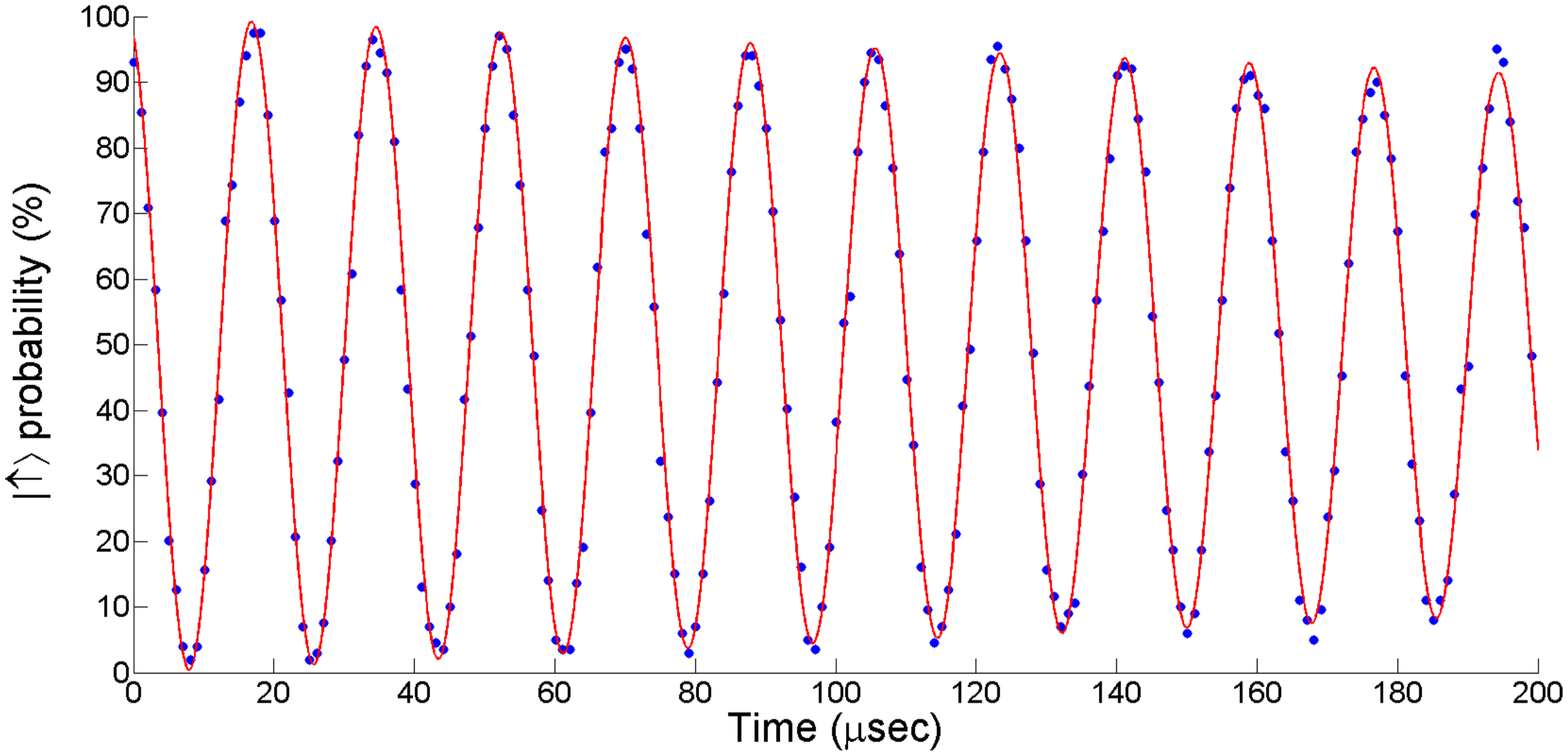}}
    \caption{
 Magnetic dipole coupling between the two states of a Zeeman qubit in a single trapped $^{88}$Sr$^+$ ion. (a) A diagram of the Zeeman qubit levels. (b) A Rabi-nutation curve. The probability of finding the ion-qubit in the $|\uparrow\rangle$ state oscillates vs. time.
    }\label{figure6}
\end{figure}

Zeeman or hyperfine qubits have a separation between
the their levels that is in the RF or microwave range. The two qubit states represent
different directions or magnitudes of a magnetic dipole and can
therefore be coupled by inducing magnetic dipole transitions. The
time-varying potential in the presence of an oscillating magnetic
field is,
\begin{equation}
V(t) = -\bold{\hat{\mu}}\cdot\bold{B}_0\cos(\bold{k}\hat{x}-\omega t
+ \phi).\label{Vmagnetic}
\end{equation}
The magnetic moment of the electron is $\bold{\hat{\mu}}= \mu_B
(g_S\bold{\hat{S}} + g_L\bold{\hat{L}} + g_I\bold{\hat{I}})$, where
$\hat{S}$, $\hat{L}$, and $\hat{I}$ are the electronic spin,
the electronic orbital angular momentum, and nuclear spin operators,
respectively, and $g_S$, $g_L$ and $g_I$ are the corresponding
$g$-factors. The Rabi frequency here is given by $\Omega_0 =
\langle\downarrow|\hat\mu\cdot\vec{B}_0|\downarrow\rangle$. For a
Zeeman qubit $\bold{\hat{\mu}}= g_S\mu_B\hat{\sigma}$. The Rabi
frequency will maximize to $\Omega_0 = 2\pi\times28B_0$ MHz when
$\bold{B}_0$ (in mT) is perpendicular to $\bold{z}$, e.g. $\bold{B}_0 =
B_0\bold{x}$. Figure \ref{figure6} shows the probability of finding a Zeeman qubit, in a $^{88}$Sr$^+$ ion, oscillating vs. time, due to magnetic-dipole coupling.
Since for different ion qubit choices and quantization magnetic
field magnitudes the qubit separation frequency varies between a few
MHz and a few GHz, the wavelength of $\bold{B}_0$, $\lambda =
1/|\bold{k}|$ will vary between $0.01$ and $1000$ meters, where the
typical $x_0$ is between $10$ and $100$ nm. The Lamb-Dicke parameter in
this case will be negligibly small, resulting in a carrier Rabi frequency is almost nearly independent of the ion motion. The disadvantage is that sideband transitions cannot be efficiently
driven. Another disadvantage is that this way single-ion qubit
addressing is impossible. The fidelity of magnetically driven
single-qubit gates is mainly limited by the amplitude and phase noise of
the magnetic RF field at the position of the ion.

\subsection{Two-photon Raman coupling}\label{Two-photon Raman coupling}
\begin{figure}[h] 
    {\includegraphics[width=0.5\textwidth]{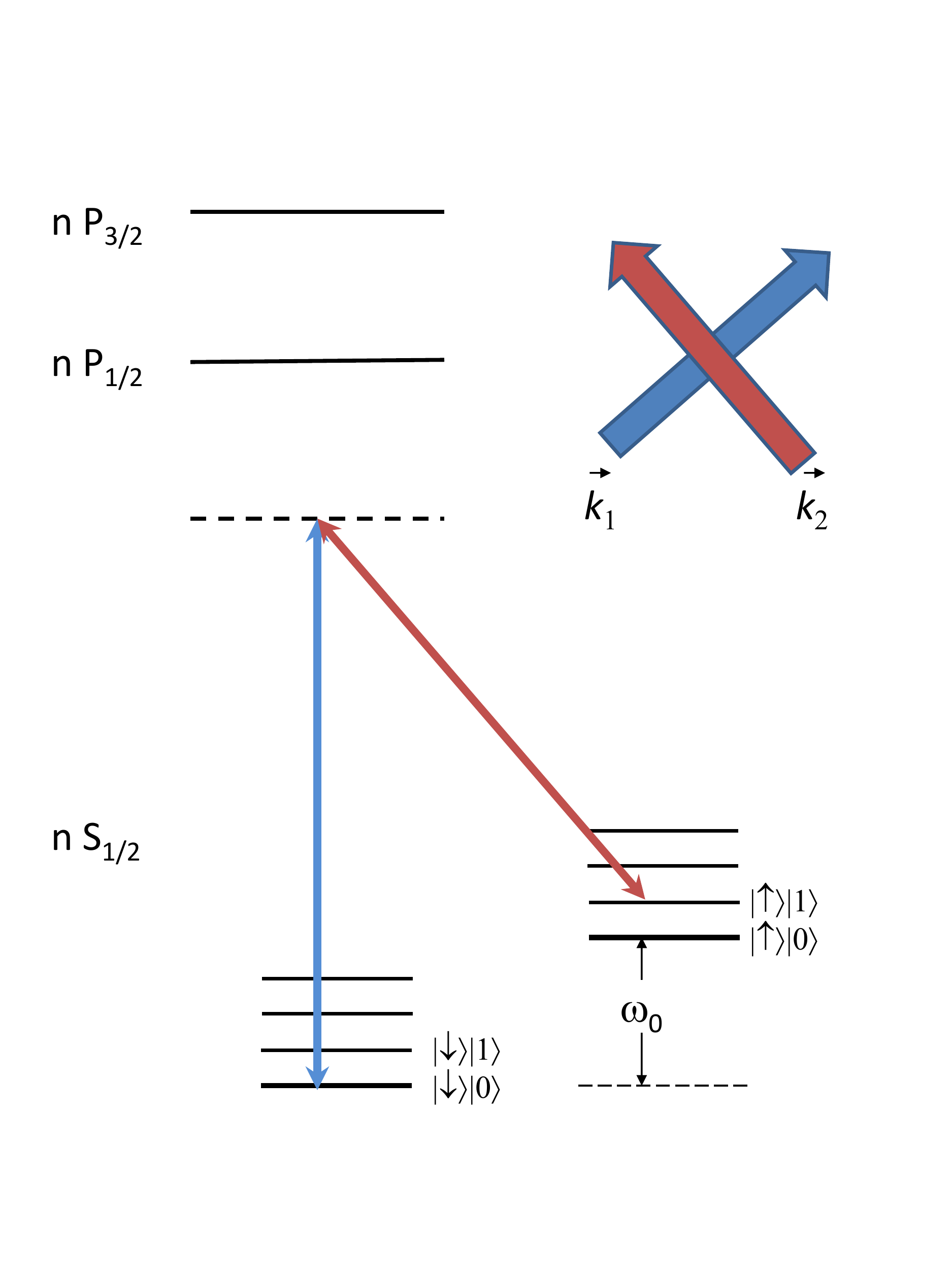}}
    \centering
    \caption{
A diagram showing Raman coupling. The Raman beams are detuned from transitions to excited states and the difference in energy between the two beams has to match the energy separation between the two qubit states. In order to excite ion motion, the two Raman beams cannot be co-propagating. A common choice is for the beams to cross at right angle with the difference in their wave-vectors parallel to the trap axis, as shown in the figure.
    }\label{figure7}
\end{figure}

Two-photon Raman transitions employ a laser as an optical
``carrier'' of the RF qubit separation frequency. In this scheme,
photons are coherently transferred between two optical modes with
frequencies that are separated by $\omega_0 + s\omega_m$ as
illustrated in Fig. (\ref{figure7}). The laser frequencies are
off-resonance, by a detuning $\Delta_i$, with optically allowed
transitions from the qubit levels to several excited levels, $e_i$.
With the qubit choices mentioned above, $e_i$ are typically the
different states in the $P_{1/2}$ and $P_{3/2}$ levels. Thus, the electric
fields of the two Raman modes can be written as,
\begin{equation}
\bold{E}_{r/b} =
\bold{\epsilon}_{r/b}E_{r/b}\cos(\bold{k}_{r/b}\cdot\hat{x}-\omega_{r/b}t+\phi_{r/b}),\label{ErbRaman}
\end{equation}
where red and blue (r/b) correspond to the lower and higher
frequency fields, respectively, and $\bold{\epsilon}$ is a unit vector
in the field polarization direction. To fully solve for the ion
evolution, the amplitudes associated with the multiple levels
involved have to be calculated. However, when the detuning is
significantly larger than the coupling strength, the amplitude
associated with the optically excited states, $e_i$, reduces as
$1/\Delta_i$ and oscillates at two time scales, $\Delta_i$, and a
slow time scale that adiabatically follows the ground-state
amplitudes. It is thus possible to ``adiabatically eliminate'' the
amplitudes of the $e_i$ states and write the equations of motion
only in terms of the ground-state amplitudes and their derivatives.
The problem thus reduces to an effective two-level problem involving
only the qubit levels. The effective Rabi frequency here is,
\begin{equation}
\Omega_0 =
\frac{E_{r}E_{b}}{4\hbar^2}\sum_{i}\frac{\langle\uparrow|\hat{\bold{d}}\cdot\bold{\epsilon_r}|e_i\rangle\langle
e_i|\hat{\bold{d}}\cdot\vec{\epsilon_b}|\downarrow\rangle}{\Delta_i}.\label{ErbRaman}
\end{equation}
The time-dependent potential would then reduce to the form of Eq.
(\ref{time dependent potential}), with $\phi=\phi_b-\phi_r$ and
$\bold{k}=\bold{k}_b-\bold{k}_r$. The carrier Hamiltonian in the RWA
will be,
\begin{equation}
\Omega_0 = \hat{H}_{int} =
\frac{\hbar\Omega_0}{2}D_{n,n}(\hat{\sigma}_+e^{i\phi} +
\hat{\sigma}_-e^{-i\phi} ) + (\Delta_{\uparrow} -
\Delta_{\downarrow})\hat{\sigma}_z.\label{H Carrier Raman}
\end{equation}
The first term on the r.h.s is identical to Eq. (\ref{Hcarrier}).
The second term on the r.h.s is due to the difference in the qubit
levels' light shifts induced by the off-resonance Raman beams,
\begin{equation}
\Delta_{\downarrow/\uparrow} = \frac{|E_r|^2}{4\hbar^2}\sum_{i}
\frac{|\langle \downarrow/\uparrow |
\hat{\bold{d}}\cdot\vec{\epsilon}_r |e_i\rangle |^2}{\Delta_{i,r}} +
\frac{|E_b|^2}{4\hbar^2}\sum_{i} \frac{|\langle \downarrow/\uparrow
| \hat{\bold{d}}\cdot\vec{\epsilon}_b |e_i\rangle
|^2}{\Delta_{i,b}}.\label{Raman Stark}
\end{equation}
The differential light shift can be tuned to zero by carefully
choosing the beam polarization and detuning.

The Lamb-Dicke parameter and therefore the ability to impart
momentum and the sensitivity of the Rabi frequency to the ions'
motion depends on the angle between the Raman beams. In order to
efficiently drive sideband transitions, the angle between the Raman
beams needs to be as large as possible and the difference in their
wave vector should be such that $\bold{k}$ has a maximal projection
along the trap axis. For the purpose of driving single qubit
rotations it is desirable to avoid changes in the Rabi frequency owing
to ion motion and therefore often a co-propagating Raman beam
configuration is chosen. Raman beams can be focused to a size
limited by the (optical) diffraction limit, typically of the order
of a wavelength, and therefore single-ion qubit addressing is in
principle possible.

Similarly to the magnetic dipole transition case, errors are caused
by classical noises. Laser intensity noise and beam pointing
fluctuations result in noise in the Rabi frequency. Phase and frequency
noise between the Raman frequency difference and the ion qubit
frequency separation also reduces the gate's fidelity. However, in this
case, in addition to classical noises, the quantum nature of light
will further contribute to the gate error \cite{Ozeri2007}. For laser beams the
effect of quantum noise on the interaction with an atom is attributed to spontaneous scattering of photons. Ground-state coherence is only affected by Raman spontaneous scattering events in which
a population is transferred between different ground-state levels \cite{Ozeri2005}, or when the scattering amplitude from the two qubit states are different \cite{Uys2010}. The
spontaneous Raman scattering probability during a single qubit gate
contributes to the gate error. This contribution reduces for larger
$\Delta_i$ values.

\subsection{Optical quadrupole coupling}\label{Optical quadrupole coupling}
\begin{figure}[h] 
   {\includegraphics[width=0.5\textwidth]{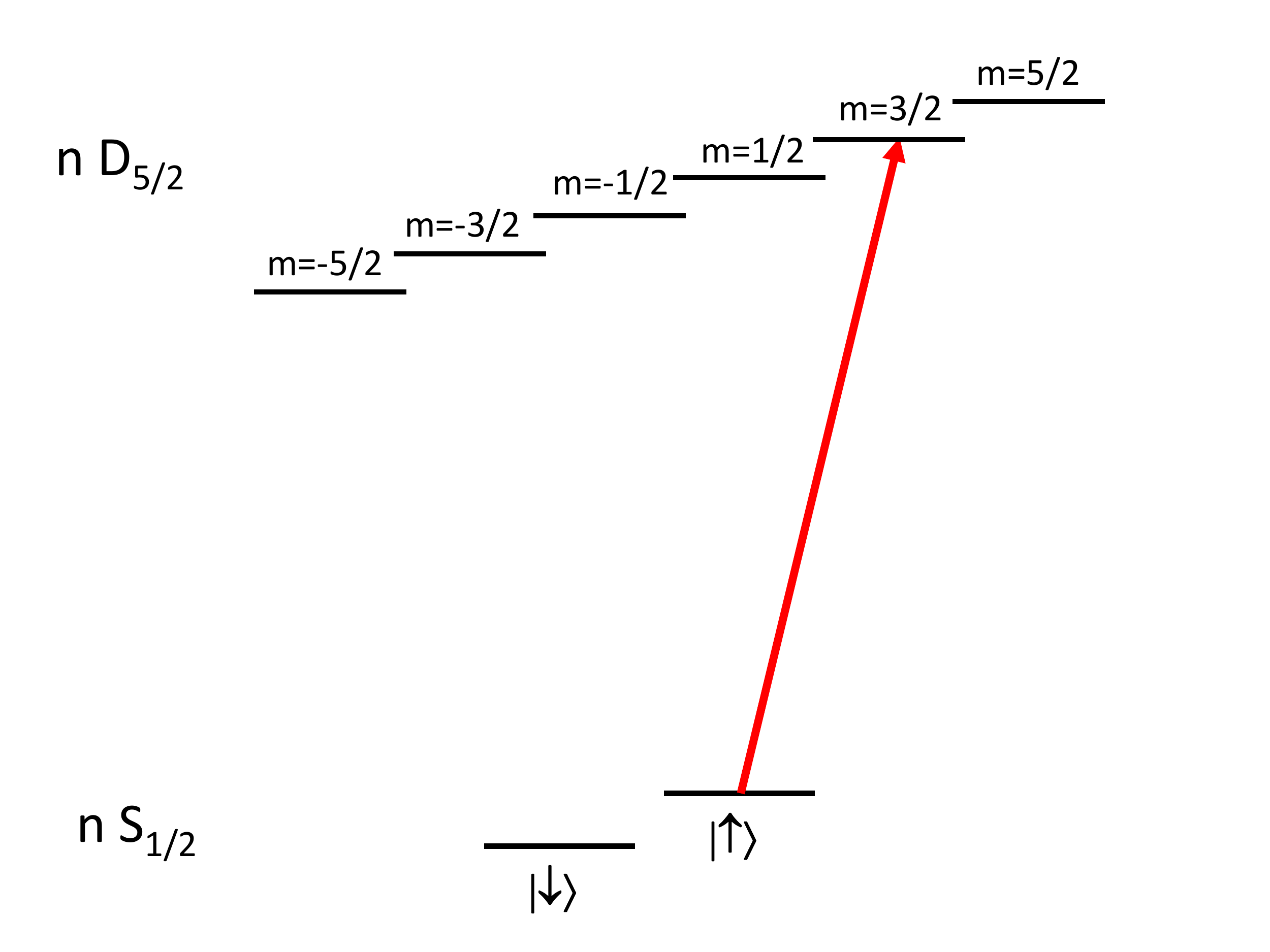}}
   {\includegraphics[width=0.5\textwidth]{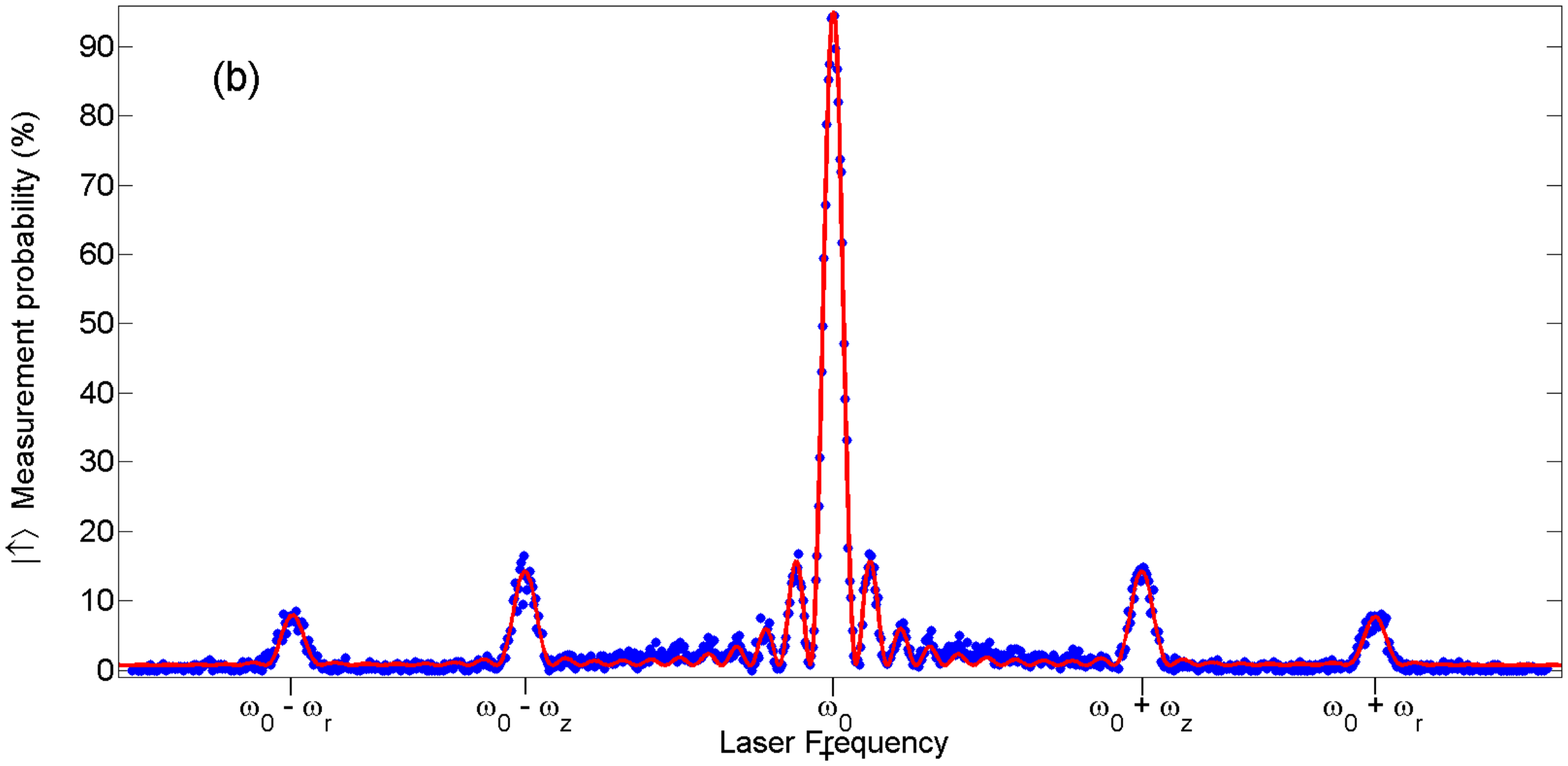}}
    \caption{
Optical electric-quadrupole coupling between the two states of an optical qubit in a single trapped $^{88}$Sr$^+$ ion. (a) A diagram of the Optical qubit levels. (b) A spectral response of the ion to the laser frequency. On both sides of the transition carrier are red and blue sidebands corresponding to emission and absorption of longitudinal (z) or radial (r) quanta of motion.
    }\label{figure8}
\end{figure}

Optical qubits are manipulated with optical fields.
Consider an optical ion qubit encoded in superpositions of the
$|S_{1/2}, m\rangle$ and $|D_{5/2},m'\rangle$ states. Since the two
qubit states have the same parity, it is only electric quadrupole
transitions, which require an electric field gradient across the
ion, which couple the two levels. Given an optical electric field,
\begin{equation}
\bold{E} = \bold{\epsilon}E_{0}\cos(\bold{k}\cdot\hat{x}-\omega_0
t+\phi),\label{E quanruple}
\end{equation}
the electric quadrupole Rabi frequency is given by,
\begin{equation}
\Omega_0 = \frac{eE_0}{2\hbar}\langle S_{1/2},m|(\bold{\epsilon\cdot
r)(k\cdot r)}|D_{5/2},m'\rangle .\label{Rabi Quadrupole}
\end{equation}
Figure \ref{figure8} shows response spectrum of an optical qubit in a single $^{88}$Sr$^+$ ion to a narrow line width laser scanning across its transition frequency. Red and Blue side-bands are apparent on both sides of the carrier transition.

Unlike two-photon Raman transitions, here the Lamb-Dicke parameter
cannot be tuned to zero. Errors in optical quadrupole single qubit
gates will result from classical noises in the intensity and
frequency of the driving laser, and in addition, from a thermal
component in the ions' motion. Another error results from the finite
probability of the $D_{5/2}$ level to decay during the gate. However,
since the lifetime of the $D$ levels is typically $\simeq 1$ sec and
much larger than the gate time, this error is typically small.

\section{Two-qubit gates}\label{Two-qubit gates}
Thus far, we considered only single-qubit rotations. In addition to
signle-qubit gates, the standard universal gate set includes CNOT
entangling operations. CNOT gates rotate the state of a target qubit
around the $x$ axis by $180^\circ$, depending on the logical state of a second,
control, qubit. Lexicographically ordered, a general two-qubit state
is written as the four vector,
\begin{equation}
\alpha|\uparrow\uparrow\rangle + \beta|\uparrow\downarrow\rangle +
\gamma|\downarrow\uparrow\rangle +
\delta|\downarrow\downarrow\rangle= \left [ \begin{array}{c}
   \alpha\\\beta\\\gamma\\\delta
   \end{array}\right ].\label{four vector}
\end{equation}
The CNOT operation is then represented by the matrix,
\begin{equation}
\left [ \begin{array}{cccc}
   1&0&0&0\\0&1&0&0\\0&0&0&1\\0&0&1&0
   \end{array}\right ]\left [\begin{array}{c}
   \alpha\\\beta\\\gamma\\\delta
   \end{array}\right ] = \left [ \begin{array}{c}
   \alpha\\\beta\\\delta\\\gamma
   \end{array}\right ].\label{four vector}
\end{equation}
Since the roles of the target and control qubits are different,
implementing this gate on an array of trapped-ion qubits requires a
physical operation that can distinguish between the two. An entangling operation that is
symmetric with respect to both ions, appears to be
an easier operation to implement. Rotating the target qubit by a Hadamard rotation, which up to a global phase can be composed from two consecutive rotations
$\hat{H}=e^{i\pi/2}\hat{R}(0,0,\pi)\hat{R}(0,\pi/2,\pi/2)$, both before and after the gate, we obtain the operation,
\begin{equation}
 [\hat{I}_1 \otimes \hat{H}_2]\left [ \begin{array}{cccc}
   1&0&0&0\\0&1&0&0\\0&0&0&1\\0&0&1&0
   \end{array}\right ][\hat{I}_1\otimes \hat{H}_2]=
\left [ \begin{array}{cccc}
   1&0&0&0\\0&1&0&0\\0&0&1&0\\0&0&0&-1
   \end{array}\right ].\label{pi phase gate}
\end{equation}
Here, a $\pi$ phase shift is imprinted on the
$|\downarrow\downarrow\rangle$ state with respect to the three other
collective spin states. Note that since this gate and the CNOT are
connected via single-qubit rotations, they are equivalent in terms of a universal gate set. A subsequent $\pi/2$ rotation of
both qubits around the $z$ axis yields,
\begin{equation}
[\hat{R}_1(\pi/2,0,\pi/2)\otimes
\hat{R}_2(\pi/2,0,\pi/2)]\left [
\begin{array}{cccc}
   1&0&0&0\\0&1&0&0\\0&0&1&0\\0&0&0&-1
   \end{array}\right ]
= e^{-i\pi/2}\left [ \begin{array}{cccc}
   1&0&0&0\\0&i&0&0\\0&0&i&0\\0&0&0&1
   \end{array}\right ].\label{pi/2 phase gate}
\end{equation}
Here, a $\pi/2$ phase is imprinted on anti-parallel spin states with
respect to parallel spin states.

Both examples above are entangling operations that rely on
imprinting a phase on certain collective spin states. It is
important to emphasize that whether a certain entangling operation
is a phase gate or not depends on the choice of basis in which it is
presented. As an example, the last operation, written in the $x$
basis is,
\begin{eqnarray}
&&[\hat{R}_1(0,\pi/2,\pi/2)\otimes \hat{R}_2(0,\pi/2,\pi/2)]\left [
\begin{array}{cccc}
   1&0&0&0\\0&i&0&0\\0&0&i&0\\0&0&0&1
   \end{array}\right ]\\ \nonumber
   && [\hat{R}_1(0,\pi/2,-\pi/2)\otimes \hat{R}_2(0,\pi/2,-\pi/2)]= \frac{e^{i\pi/4}}{\sqrt{2}}
\left [ \begin{array}{cccc}
   1&0&0&-i\\0&1&-i&0\\0&-i&1&0\\-i&0&0&1
   \end{array}\right ]\equiv \hat{U},\label{SM gate}
\end{eqnarray}
a collective spin flip operation,
\begin{eqnarray}
\hat{U}|\downarrow\downarrow\rangle = \frac{e^{-i\pi/4}}{\sqrt{2}}\left (
|\downarrow\downarrow\rangle +i|\uparrow\uparrow\rangle   \right ),\\
\hat{U}^2|\downarrow\downarrow\rangle = |\uparrow\uparrow\rangle.
\end{eqnarray}

In fact, most of the proposed implementations of an entangling gate
for trapped-ion qubits can be thought of as a phase gate in one basis and a collective spin-flip operation in another.

\subsection{Quantum phases in cyclic motion}\label{Quantum phases in cyclic motion}
Entangling phase gates change certain two-qubit collective spin
states only by phase factor multiplication. One of
those collective spin states undergoes cyclic evolution during the gate, at the end of which, it returns to its initial
state with an acquired phase. The phase factors that accompany the
cyclic evolution of a quantum state were first studied by Berry for
adiabatic state evolution \cite{Berry1984} and later by Aharonov and
Anandan for the general case \cite{AharonovAnandan1987}. Following
\cite{AharonovAnandan1987}, let us consider a general quantum state
undergoing a cyclic evolution with a period $\tau$,
\begin{equation}
|\Psi(\tau)\rangle = e^{i\phi}|\Psi(0)\rangle.\label{cyclic
evolution}
\end{equation}
The state evolves according to the time-dependent
Schr\"{o}dinger equation,
\begin{equation}
\hat{H}(t)|\Psi(t)\rangle = i\hbar(d/dt)|\Psi(t)\rangle.\label{time
dependent Sch.}
\end{equation}
We define the ``bare'' state,
\begin{equation}
|\Phi(t)\rangle = e^{-if(t)}|\Psi(t)\rangle.\label{bare state}
\end{equation}
such that $f(\tau)-f(o) = \phi$. We can now write a differential
equation for $f$,
\begin{equation}
\frac{df}{dt}=-\frac{1}{\hbar}\langle\Psi(t)|\hat{H}(t)|\Psi(t)\rangle+i\langle\Phi(t)|\frac{d}{dt}|\Phi(t)\rangle.\label{Equation
f}
\end{equation}
The phase accumulated by $|\Psi\rangle$ at time $\tau$ is given by
integrating Eq.(\ref{Equation f}),
\begin{equation}
\phi =
-\frac{1}{\hbar}\int_{0}^{\tau}\langle\Psi(t)|\hat{H}(t)|\Psi(t)\rangle
dt +i\int_{0}^{\tau}\langle\Phi(t)|\frac{d}{dt}|\Phi(t)\rangle dt
\equiv \phi_D + \gamma.\label{total phase}
\end{equation}
There are two different contributions to the total phase. The first
term on the r.h.s in Eq.(\ref{total phase}) is $\phi_D$, the Dynamic
phase, i.e., the time integration of the energy expectation value
divided by $\hbar$. The second term on the r.h.s is the geometric
part of the phase, $\gamma$. This part is independent of the exact
Hamiltonian under which the state evolves but rather on its
evolution path in parameter space. Assuming that at every time
instance the wavefunction can be mapped onto a parameter space
$\alpha$, $|\Phi(t)\rangle \equiv |\Phi_\alpha\rangle$, then $\gamma$
can be written as an integral along a closed $\alpha$-contour,
\begin{equation}
 \gamma = i\int_0^\tau\langle \Phi(t) |
\frac{d}{dt}|\Phi(t)\rangle dt = i\oint_c \langle \Phi_\alpha |
\vec{\triangledown}_\alpha |\Phi_\alpha\rangle \cdot
d\vec{\alpha}.\label{geometric phase}
\end{equation}

\begin{figure}[h] 
   {\includegraphics[width=0.5\textwidth]{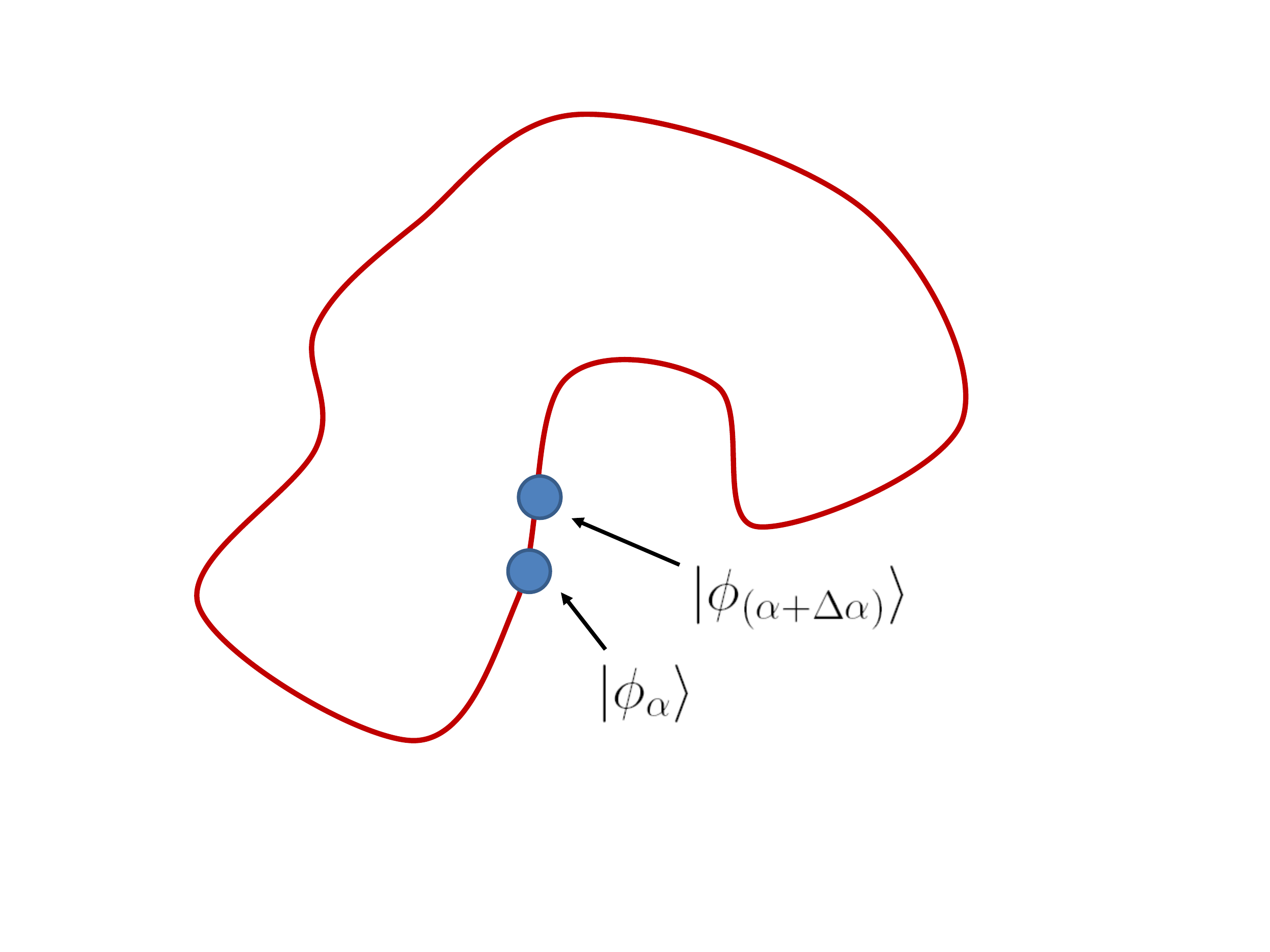}}
   \centering
   \caption{
A wavefunction evolving around a closed contour in parameter space. The two circles represent the wavefunction at $\alpha$ and $\alpha+\Delta\alpha$.
    }\label{figure9}
\end{figure}

To illustrate the geometric origin of $\gamma$, consider a
wavefunction evolving around the closed contour in parameter space,
as shown in Fig.(\ref{figure9}). The infinitesimal phase difference,
$\Delta\gamma$ between $|\Phi_\alpha\rangle$ and
$|\Phi_{\alpha+\Delta\alpha}\rangle$ is defined as,
\begin{equation}
 e^{-i\Delta\gamma} = \frac{\langle\Phi_\alpha|\Phi_{\alpha+\Delta\alpha}\rangle}{|\langle\Phi_\alpha|\Phi_{\alpha+\Delta\alpha}\rangle|}.\label{exp geometric phase diff}
\end{equation}
Taking the log of both sides of Eq.(\ref{exp geometric phase diff})
yields,
\begin{equation}
-i\Delta\gamma \simeq \langle \Phi_\alpha |
\vec{\triangledown}_\alpha |\Phi_\alpha\rangle \cdot
d\vec{\alpha}.\label{geometric phase diff}
\end{equation}
The geometric phase is thus calculated by integrating
Eq.(\ref{geometric phase diff}) around the $\alpha$-contour yielding
Eq.(\ref{geometric phase}). By defining the vector potential
\begin{equation}
\bold{A}_\alpha =  i\langle \phi_\alpha | \vec{\triangledown}_\alpha
|\phi_\alpha\rangle,\label{vector potential}
\end{equation}
we can write an effective ``magnetic field'',
\begin{equation}
\bold{B}_\alpha =
\vec{\triangledown}\times\bold{A}_\alpha.\label{magnetic field}
\end{equation}
Using Stokes theorem we then express the geometric phase as the
effective magnetic field flux through the parameter space area, $S$,
\begin{equation}
\gamma = - \int \int_S \textbf{B} \cdot \textbf{n} dS.\label{Stokes}
\end{equation}
Here $\textbf{n}$ is the unit vector normal to $S$. The encirclement
of a finite area in parameter space is therefore a necessary
condition for accumulating a geometrical phase in cyclic
motion.

An ion qubit is a (pseudo-) spin connected to a harmonic oscillator.
A phase gate such as (\ref{pi/2 phase gate}) can be implemented by
acquiring either a dynamic or a geometric phase or both, during a
cyclic motion either in spin configuration space or in the h.o.
phase space. Cirac and Zoller's seminal 1995 proposal for an entangling gate with trapped-ion qubits \cite{CiracZoller1995} implements a phase gate where the
phase is geometric and the cyclic motion is in spin configuration
space. A later proposal by the same authors uses a dynamic phase
acquired during a cyclic motion in phase space \cite{CiracZoller2000}. In the following section we
will concentrate on a class of gates, initially proposed by S\o rensen
and M\o lmer \cite{SM1999} and independently by Milburn and co-authors \cite{Milburn2000}, which uses spin-dependent
forces to acquire a sum of geometric and dynamic phases during a
cyclic motion in h.o. phase space.

\subsection{Driven harmonic oscillator}\label{Driven harmonic oscillator}
A classical harmonic oscillator evolves periodically in phase space
under the influence of a force, $F(t)=F_0\cos(\omega t)$,
oscillating at a frequency slightly off-resonance from its harmonic
frequency, $\omega = \omega_m-\delta$. Starting at rest,
$x(t=0)=p(t=0)=0$, the position and momentum of the oscillator at
subsequent times are given by,
\begin{eqnarray}
 && x(t) = \frac{F_0/m}{\omega^2-\omega_m^2}\left [ \cos(\omega t) -  \cos(\omega_m t) \right ]\\
 && p(t) = \frac{F_0}{\omega^2-\omega_m^2}\left [ \omega_m \sin(\omega_m t) -  \omega \sin(\omega t) \right ].\label{classical driven osc.}
\end{eqnarray}
Initially the driving force increases the oscillation amplitude
until, at time $t=\pi/\delta$, the force and the oscillator are
out of phase. From this time onward the force dampens the oscillator
until, at time $t=2\pi/\delta$, the oscillator returns to its
initial state. To make an analogy with the quantum case, we
normalize the h.o. position by $2x_0$ and the momentum by $p_0
\equiv \hbar/x_0$ and move to a frame rotating at the oscillator
frequency. Neglecting terms that are proportional to
$\delta/\omega_m$ we arrive at,
\begin{eqnarray}
 && x'(t) = \frac{x(t)}{2x_0}\cos(\omega_mt)-\frac{p(t)}{p_0}\sin(\omega_mt) \equiv \frac{F_0x_0}{2\hbar\delta}\left [1 - \cos(\delta t) \right ]\\
 && p'(t) = \frac{x(t)}{2x_0}\sin(\omega_mt)+\frac{p(t)}{p_0}\cos(\omega_mt) \equiv \frac{F_0x_0}{2\hbar\delta}\left [\sin(\delta t) \right ].\label{classical driven osc. rotating frame}
\end{eqnarray}
We define the complex function $\alpha (t) = x'(t) + ip'(t)$. The
h.o. oscillator motion in the complex plane is along the curve,
\begin{equation}
\alpha (t) = \frac{F_0x_0}{2\hbar\delta}\left [1 - e^{i\delta t}
\right ].
\end{equation}
The h.o. therefore periodically evolves around a circle in the
complex plane, the area of which is
\begin{equation}
S =  \pi\big[\frac{F_0x_0}{2\hbar\delta} \big
]^2.\label{phase space area}
\end{equation}

\begin{figure}[h] 
   {\includegraphics[width=0.5\textwidth]{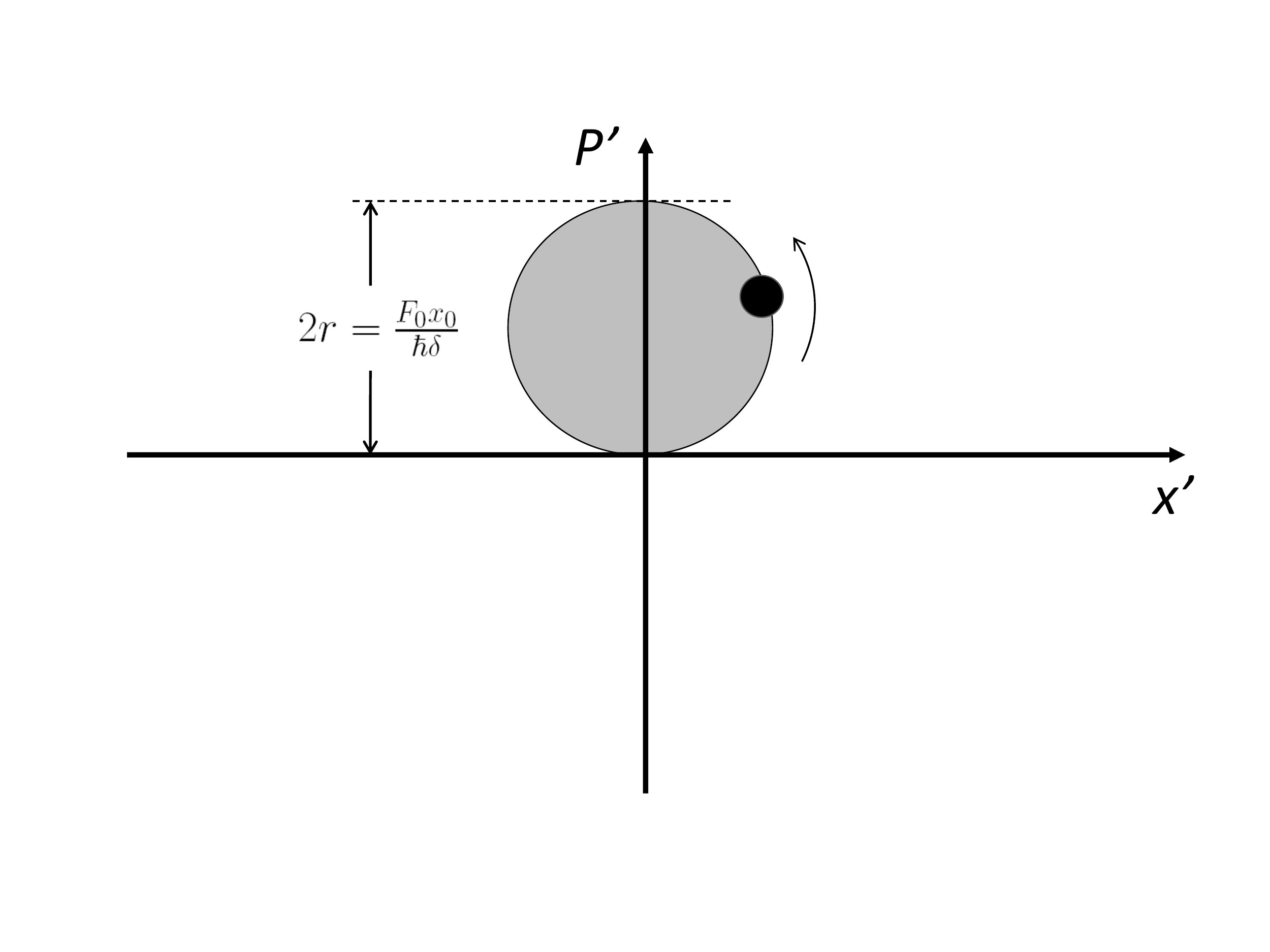}}
   \centering
   \caption{
An illustration of the quantum driven harmonic-oscillator wavefunction trajectory through phase-space. The oscillator is startiong from the ground-state and is off-resonantly driven with a periodic force with amplitude $F_0$. The area that the oscillator is circulating in phase space is equal to $S =  \pi\big[\frac{F_0x_0}{2\hbar\delta} \big
]^2$.
    }\label{figure10}
\end{figure}

We now turn to solving the quantum, driven, harmonic oscillator. The
quantum Hamiltonian of a driven h.o. is,
\begin{equation}
H = \hbar\omega_m (\hat{a}^\dagger\hat{a} + 1/2) + F(t)\hat{x}\equiv
H_0 + V(t). \label{Driven h.o. Hamiltonian}
\end{equation}
Note that here we assume that the force is spatially uniform and
therefore it does not vary across the h.o. wavefunction. Moving to the
interaction representation and using the RWA we write,
$i\hbar\frac{d}{dt}|\Psi\rangle_{I} = V_{I}|\Psi\rangle_{I}$, with,
\begin{equation}
V_{I} = \frac{F_0x_0}{2}\left ( \hat{a}^\dagger e^{i\delta t} +
\hat{a} e^{-i\delta t} \right ). \label{Driven h.o. Hamiltonian}
\end{equation}
$V_I$ does not commute with itself at different times; however, for a
short enough $\Delta t$, we can still write,
\begin{equation}
|\Psi(t+\Delta t) \rangle_I = e^{-\frac{i}{\hbar}V_I\Delta
t}|\Psi(t)\rangle_I =
e^{(\Delta\alpha\hat{a}^\dagger+\Delta\alpha^\star\hat{a})}|\Psi(t)\rangle_I
\equiv \hat{D}(\Delta\alpha)|\Psi(t)\rangle_I, \label{int
propagator}
\end{equation}
where $\hat{D}(\Delta\alpha)$ is the displacement operator,
\begin{equation}
\hat{D}(\alpha)|0\rangle = |\alpha\rangle =
e^{-\frac{1}{2}|\alpha|^2}\sum_{n=0}^{\infty}\frac{\alpha^n}{\sqrt{n!}}|n\rangle.
\label{displacement operator}
\end{equation}
When operating on the h.o. ground-state the displacement operator
forms coherent states of motion. Here the infinitesimal displacement
is,
\begin{equation}
\Delta\alpha(t) = -\frac{i}{2\hbar}F_0x_0e^{i\delta t}\Delta t
\equiv [\Delta x/2x_0 + i p/p_0]. \label{delta alpha}
\end{equation}
The real and imaginary part of the displacement can be identified
with its position and momentum parts, respectively. To add two
infinitesimal displacements, we use the Baker-Hausdorff formula,
\begin{equation}
e^{\hat{A}}e^{\hat{B}}=
e^{\hat{A}+\hat{B}}e^{\frac{1}{2}[\hat{A},\hat{B}]}, \label{Baker
Hausdorff}
\end{equation}
for operators $\hat{A}$ and $\hat{B}$ that commute with their
commutator,
$[\hat{A},[\hat{A},\hat{B}]]=[\hat{B},[\hat{A},\hat{B}]]=0$. We thus
get,
\begin{equation}
\hat{D}(\alpha)\hat{D}(\beta) = \hat{D}(\alpha + \beta)e^{i
Im(\alpha\beta^\star)}. \label{adding two displacements}
\end{equation}
The displacement operator can accordingly be re-written as,
\begin{equation}
\hat{U}(t=0,t)=
\hat{D}(\sum_{i=1}^N\Delta\alpha_i)e^{iIm[\sum_{j=2}^N\Delta\alpha_j(\sum_{k=1}^{j-1}\Delta\alpha_k)^\star]}=
\hat{D}(\alpha)e^{iIm\oint \alpha^\star d\alpha},\label{displacement
operator alpha}
\end{equation}
with, $\alpha = \sum_{i=1}^N\Delta\alpha_i$. The total displacement
at time, $t$, is therefore,
\begin{equation}
\alpha(t) = \int_o^{t}\Delta\alpha(t)dt =
-\frac{i}{2\hbar}\int_o^{t}F_0x_0e^{i\delta t'}dt' =
\frac{F_0x_0}{2\hbar\delta}(1-e^{i\delta t}).\label{alpha}
\end{equation}
The quantum wavefunction and the classical h.o. therefore follow an
\emph{identical} trajectory in phase space, illustrated in Fig.
\ref{figure10}. As in the classical case, after time $\tau_g =
2\pi/\delta$, the h.o. wavefunction returns to its initial state.

The total phase that is accumulated by the end of the gate is,
\begin{equation}
\phi = Im\oint \alpha^\star d\alpha=
\frac{\pi}{2}\big{(}\frac{F_0x_0}{\hbar\delta}\big{)}^2,\label{h.o.
total phase}
\end{equation}
To implement the phase gate (\ref{pi/2 phase gate}), the force
magnitude and detuning are adjusted such that $\phi=\pi/2$.
Comparing the phase to the area encircled in phase space during the
gate, Eq. (\ref{phase space area}), we obtain,
\begin{equation}
\gamma = 2S.\label{geometric phase vs. phase space area}
\end{equation}
A proportionality between the phase accumulated per cycle and the
area encircled by the wavefunction in phase space suggests that the phase has a
geometric origin.

To make the connection with Eq.(\ref{total phase}) we calculate the
dynamic and geometric contributions to $\phi$. The dynamic phase is
simply,
\begin{eqnarray}
&& \phi_D =
-\frac{1}{\hbar}\int_0^{\tau_g}\langle\alpha(t)|\hat{V}_I(t)|\alpha(t)\rangle
dt \\ \nonumber &&= \frac{F_0x_0}{2\hbar}\int_0^{2\pi/\delta}\big (
\alpha^{\star}e^{i\delta t} +  \alpha^{-i\delta t} \big ) = \pi
\big{(}\frac{F_0x_0}{\hbar\delta}\big{)}^2 = 2\phi.\label{dynamic
contribution to the phase}
\end{eqnarray}
The geometric part of the phase is,
\begin{equation}
 \gamma = i\oint_c \langle \alpha |
\vec{\triangledown} |\alpha\rangle \cdot
d\vec{\alpha}.\label{geometric contribution ot the phase}
\end{equation}
Here phase space is the parameter space (a complex plane) and the
wavefunction trajectory is parameterized by the complex displacement
$\alpha$. We evaluate the integrand along the wavefunction path by,
\begin{equation}
 \langle \alpha | \vec{\triangledown} |\alpha\rangle \cdot d\vec{\alpha} = \langle\alpha|\alpha +
\Delta\alpha\rangle - 1.\label{path integrand}
\end{equation}
Coherent states are an over-complete basis of wavefunctions and are
therefore not mutually orthogonal. The overlap between two coherent
states $\alpha$ and $\beta$ is,
\begin{equation}
 \langle\alpha|\beta\rangle =
e^{-\frac{1}{2}(|\alpha|^2+|\beta|^2)+\alpha^\star\beta}.\label{coherent
state overlap}
\end{equation}
We therefore arrive at,
\begin{equation}
 \langle\alpha|\alpha+\Delta\alpha\rangle -1 \simeq
\frac{1}{2}(\alpha^\star\Delta\alpha - \alpha\Delta\alpha^\star) =
iIm(\alpha^\star\Delta\alpha).\label{path integrand explicit}
\end{equation}
leading to the geometric phase,
\begin{equation}
\gamma = -Im\oint \alpha^\star d\alpha= -\phi. \label{h.o. geometric
phase}
\end{equation}
The total phase acquired during the gate is consequently the sum of
partially canceling dynamic and geometric contributions. Although of
different origin, the geometric and dynamic parts of the phase are
proportional to each other. The total phase, therefore, is also
proportional to the geometric phase and accordingly, to the area
encircled in phase space \cite{Zhu2003}. The fact that $\phi$ depends on the
wavefunction path in phase-space rather than on dynamic properties of
the gate Hamiltonian contributes to the phase gate robustness
against noise.

Following Eq.(\ref{vector potential}) we can define a vector potential,
\begin{equation}
\vec{A} = i\langle \alpha | \vec{\triangledown} |\alpha\rangle =
\left [  -Im(\alpha), +Re(\alpha), 0 \right ].\label{ho vector
potential}
\end{equation}
The effective magnetic field (as in Eq.(\ref{magnetic field})) is then
perpendicular to the complex plane, i.e. $\bold{B}=B_z\bold{z}$ and
constant,
\begin{equation}
B_z = \frac{\partial Re(\alpha)}{\partial Re(\alpha)} + \frac{\partial
Im(\alpha)}{\partial Im(\alpha)} = 2.\label{effective B phase space}
\end{equation}
By using the Stokes theorem, Eq. (\ref{Stokes}), we again find that the geometric phase equals
minus twice the phase-space area encircled, as in equations (\ref{geometric
phase vs. phase space area}) and (\ref{h.o. geometric
phase}). The motion of the h.o. wavefunction in
phase-space (in the interaction representation) is therefore
analogous to the motion of a charged particle in a uniform magnetic
field. The acquired geometric phase is therefore analogous to the
Aharonov-Bohm phase.

\subsection{Spin-dependent forces}\label{Spin-dependent forces}
Acquiring a phase is not sufficient realizing of a
two-qubit phase gate such as in  Eq. (\ref{pi/2 phase gate}). The phase
acquired and therefore also the wavefunction trajectory through
phase-space have to be dependent on the collective two-qubit spin
state. To this end, spin-dependent forces have to be used.

Since light couples differently to different spin states, it can be used
to apply spin-dependent forces on trapped-ion qubits. Spin-dependent
forces can be applied to Zeeman or hyperfine ion qubits by using the
differential light shift, Eq. (\ref{Raman Stark}). Here, a
spatial variation in the light shift induces different forces on
ions in different collective spin states. Ions in a collective spin
superposition move along position-momentum paths that are entangled
with the different spin states. Since the force evolution is cyclic,
spin and motion are disentangled at the end of the gate. Here,
since the different collective spin states are eigenstates of
$\hat{\sigma}_z\otimes\hat{\sigma}_z$, the gate is referred to as a
$\sigma_z$ gate.

It is somewhat less trivial using the light shift, due to
coupling to a third level, to apply spin-dependent forces to an
optical ion qubit. This is because for a given transition
wavelength, when one of the qubit levels is connected to a third level,
only that level will be light shifted. Forces on the two ion qubits
will only depend on one of the qubit states. Another example of an
ion qubit on which such $\sigma_z$ gates fail are ``magnetic field
insensitive'' qubits. Here, since the magnetic field couples
primarily to the electron spin, and to a far lesser degree to
the nuclear spin, ``magnetic field insensitive'' qubits have equal
amplitudes of different electron spin contributions to the two qubit
states \cite{Lee2005}. Light also couples primarily to the electron spin and
therefore the difference in light shifts is very small and is mainly
due to the small difference in detuning between both qubit states. These
small differential light shifts are insufficient to efficiently
drive an entangling gate. Entangling phase gates on the two
ion qubits mentioned above are possible, however, when differential
forces are applied on spin states that lie on the Bloch sphere
equatorial plane. In the measurement basis, the action of these gates
is similar to Eq.(\ref{SM gate}). Since this gate is a phase gate in
the basis of eigenstates of
$\hat{\sigma}_\phi\otimes\hat{\sigma}_\phi$, where
$\hat{\sigma}_\phi=\cos{\phi}\hat{\sigma}_x +
\sin{\phi}\hat{\sigma}_y$, i.e., $\phi$ is the angle form the
positive $x$-axis, the gate is often referred to as a $\sigma_\phi$
gate.

\subsection{$\sigma_z$ gates}\label{sigma z gates}
In the case of magnetic field-sensitive Zeeman or hyperfine qubits,
$\sigma_z$ gates can be driven with a pair of beams similar to the
single-qubit Raman gate case, Eq.(\ref{ErbRaman}). Here, however, the
difference in frequency between the beams is far from the qubit
transition frequency, $\omega_0$, and is close to resonance with the
trap motional frequency, $\omega_b - \omega_r = \omega_m + \delta$.
In the RWA, no terms are resonant and therefore no direct coupling
between the two qubit levels occurs. Instead, the only effect of the
light field is to shift the two qubit levels by
$\Delta_{\uparrow/\downarrow}$, given in Eq. \ref{Raman Stark}.

To form a spatially varying potential, the two beams are crossed such
that the vector difference of their wavevectors, $\Delta k = k_b -
k_r$, is parallel to the axial direction of the trap. By choosing
beam polarizations mutually perpendicular and with a magnetic
field direction that is perpendicular to both one obtains a local polarization of the light field that varies periodically between
right circular, linear, left circular, a linear polarization
perpendicular to the previous, and back to right circular, over a
period of $2\pi/\Delta k$. This configuration is often referred to
as the lin $\perp$ lin configuration, known from Sisyphus cooling.
The light potential of each of the qubit levels can hence be written
as,
\begin{equation}
\Delta_{\downarrow/\uparrow}(x,t) =
\frac{1}{2}(\Delta_{{\downarrow/\uparrow},+} +
\Delta_{{\downarrow/\uparrow},-}) +
\frac{1}{2}\big{[}\Delta_{{\downarrow/\uparrow},+}-\Delta_{{\downarrow/\uparrow},-}\big{]}\cos(\Delta\vec{k}\cdot\vec{x}-\Delta\omega
t+\Delta\phi).\label{Gate Stark potential}
\end{equation}
Here $\Delta_{{\downarrow/\uparrow},+}$ and
$\Delta_{{\downarrow/\uparrow},-}$ are the light shifts of the
$\downarrow$ and $\uparrow$ qubit levels owing to the presence of a
right and left circularly polarized field respectively. The force on
each of the qubit levels is therefore,
\begin{equation}
F_{\downarrow/\uparrow}(x,t) =
-\frac{d\Delta_{\downarrow/\uparrow}(x,t)}{dx} = \frac{\hbar
k}{2}\big{[}\Delta_{{\downarrow/\uparrow},+}-\Delta_{{\downarrow/\uparrow},-}\big{]}\cos(\Delta
kx-\Delta\omega t+\Delta\phi)\label{Gate force}
\end{equation}

With two equal-mass ions in the trap, there are two normal modes of
motion along the trap longitudinal axis. The center of mass (CM)
mode, where the two ions oscillate in phase, has an angular
frequency that equals that of a single ion, $\omega_{cm}=\omega_m$.
The second mode is the ``stretch'' (ST) mode, where the two ions
oscillate out of phase, has an angular frequency of
$\omega_{st}=\sqrt{3}\omega_m$ \cite{DFV James}. The position
operator of each ion can be written as,
\begin{equation}
\hat{x}_i = x_{eq,i} + \frac{x_{cm0}}{\sqrt{2}}(\hat{a}^\dagger_{cm}
+ \hat{a}_{cm}) \pm \frac{x_{st0}}{\sqrt{2}}(\hat{a}^\dagger_{st} +
\hat{a}_{st}),
\end{equation}
where $\hat{a}^\dagger_{cm}$ ($\hat{a}_{cm}$) and
$\hat{a}^\dagger_{st}$ ($\hat{a}_{st}$) are the creation
(annihilation) operators of the CM and ST modes, respectively. The
root mean square of the ground-states' spatial spread of the two
normal modes are $x_{cm0}$ and $x_{st0}$. The $+$ ($-$) signs
correspond to the ion with a positive (negative) equilibrium $x$
position. This sign change is due to the ions' opposite direction of
motion in the ST mode. When exciting the CM mode the total force is
the sum of forces on the two ions, and when exciting the ST mode the
total force is the difference in forces. The treatment of a driven
h.o. above assumes a uniform force. In the Lamb-Dicke regime we
neglect the variation of the force \ref{Gate force} over the ions'
wavefunction spread, thus neglecting wavefunction squeezing effects
during the gate \cite{Gate squeezing Oxford}. The force is
accordingly considered as uniform with a value given by Eq.(\ref{Gate
force}) at the ions' equilibrium position.

\subsection{$\sigma_\phi$ gates}\label{sigma phi gates}
As discussed in section \ref{Spin-dependent forces}, it is hard to apply spin-dependent forces to certain ion qubit realizations such as optical-qubits or hyperfine clock-transition qubits \cite{Lee2005}. It is possible, however, to apply forces that will depend on their superpositions. In particular, it is possible to apply forces that will be equal in magnitude but opposite in direction on the two qubit states: $|+\phi\rangle = \frac{1}{\sqrt{2}}(|\uparrow\rangle + e^{i\phi}|\downarrow\rangle$ and $|-\phi\rangle = \frac{1}{\sqrt{2}}(|\uparrow\rangle + e^{-i\phi}|\downarrow\rangle$). Here, spin-dependent forces are applied in a basis that lies in the equatorial plane of the initial qubit (ion energy eigenstates) Bloch-sphere. As shown in Eq. \ref{SM gate}, in the qubit basis this operation will look like a collective spin rotation. This kind of gate was initially proposed by S\o rensen and M\o lmer \cite{SM1999} and demonstrated both on hyperfine as well as optical qubits \cite{Haljan2005,BenhelmNature2008}.

\section{Summary}\label{Summary}
In this tutorial we reviewed several of the building blocks necessary for QIP with trapped-ion qubits. We focused on ways by which a universal quantum gate set can be implemented, and in particular on quantum gates which are driven by electro-magnetic fields in the far-field region. Here, in order to impart momentum to a cold ion, optical fields are needed. Use of optical fields requires relatively narrow line-width lasers. Since atomic transitions of ions are typically in the violet to ultra violet spectral regions, lasers are needed at this wavelength range. The advent of Gallium-Nitride diode lasers has significantly lowered the cost of near-UV and violet laser systems. In fact, for some ion species considered as quantum information carriers, such as Ca$^+$ or Sr$^+$, all the necessary wavelengths for QIP operations are available either directly, or with frequency-doubled, diode laser systems. There are, however, several difficulties and disadvantages in using light fields as control fields for QIP. One difficulty is in integrating light into a large-scale ion-trap architecture \cite{Kim2009}. Realistically, all the routing (e.g. with optical fibers) and switching will have to be multiplexed into the trap array structure. This is a non-trivial task and several initial steps have been taken in this direction  \cite{Vandevender2010}. Another disadvantage is that, although treated here in a classical fashion, a light field (as well as any other control field) is essentially a quantum field, featuring quantum noise in its operation on a qubit. Quantum noise in the operation of a certain control field will pose a quantum limit to the operation's fidelity. In the operation of light fields on an atom, this error can be thought of as originating from spontaneous scattering of photons during the gate operation \cite{Ozeri2007}. Optical light fields suffer from this type of error to a much larger extent than do control fields in the radio-frequency or microwave frequency range. It is important to note that our review is not at all exhaustive. For example, in the last few years, alternative methods have been suggested, with a few initial demonstrations, in which a universal quantum gate set is performed on trapped-ion qubits using only microwave fields \cite{Wunderlich2001, Chiaverini2008, Ospelkaus2008, Wunderlich2009, Ospelkaus2011}. Here, momentum transfer to trapped-ions is achieved with strong magnetic field gradients.

As mentioned in the introduction to this tutorial, despite experimentally realizing of all the basic building blocks for a trapped-ion quantum computer, there are many different difficulties that must be overcome until a large-scale device is finally realized. One such difficulty lies in the way decoherence, or error, scales as a function of the quantum register size. Without active error-correction, the error probability in a quantum register after a given time would scale exponentially with the register size, thus negating large-scale quantum computing. Fortunately, it has been shown that by using certain error-correction protocols, and given that the error-probability per operation is below a certain threshold, large scale quantum computing is rendered fault-tolerant \cite{Aharonov1996,Preskill1998}. This means that under these conditions, the error in a quantum computation would be independent of the register size or the number of computational steps. The fault-tolerance threshold is not cast in stone and depends on many parameters such as the exact noise mechanism, the error-correction protocol used, the register architecture and so forth. However, under rather general assumptions, this threshold has been estimated to be in the $10^{-2}-10^{-4}$ range \cite{Knill2005}. In the last few years, a large effort has been directed toward perfecting all the basic operations on trapped-ion qubits, e.g., initialization, detection, memory error, and one- and two-ion gates, and reducing the error in these operations as much as possible. Currently, typical errors in the gates described above are also in the $10^{-2}-10^{-4}$ range \cite{Knill2008,Myerson2008,BenhelmNature2008,Anna2011,Brown2011}, with good prospects for further improvements.

There is a broad consensus, we believe, that a very large gap exists between the QIP capabilities demonstrated so far with trapped-ion qubits, or any other qubit technology for that matter, and realizing a large-scale quantum computer. On the other hand, this gap is also widely believed to be purely technological rather than fundamental. If this is the case, with more effort and time, a quantum computer will eventually be realized. Failure to eventually realize a quantum computer will be glamorous because it will indicate a need to modify the laws of quantum mechanics when describing large-scale systems.

\section{Acknowledgements}\label{Acknowledgements}
I would like to thank Martina Knoop and Richard Thompson for encouraging me to write this tutorial and Shlomi Kotler for helpful remarks on the manuscript and help in the preparation of Figures.


\begin{thebibliography}{}
\bibitem{Ike&Mike} M.~A.~Nielsen and I.~L.~Chuang, \emph{``Quantum Information and Quantum Information''}, Cambridge University Press, 2000
\bibitem{Schrodinger1935} E.~Schr\"{o}dinger, Die Naturwissenschaften, \textbf{23}, 844 (1935)
\bibitem{Wineland_bible} D.~J.~Wineland \emph{et. al.}, J. Res. Nat. Inst. Stand. Tech., \textbf{103}, 259 (1998)
\bibitem{LeibfriedRMP2003} D.~Leibfried, R.~Blatt, C.~Monroe, and D.~J.~Wineland, Rev. Mod. Phys., \textbf{75}, 281 (2003)
\bibitem{Home2009} J.~Home \emph{et. al.}, Science, \textbf{325}, 1227 (2009)
\bibitem{Kielpinski2002} D.~Kielpinski, C.~Monroe, and D.~J.~Wineland Nature \textbf{417}, 709 (2002)
\bibitem{Kim2009} J.~Kim and C.~Kim, Quant. Inf. and Comp., \textbf{9}, 181 (2009)
\bibitem{Deutsch1989} D.~Deutsch, Proc. R. Soc. London Ser. A \textbf{425}, 73 (1989)
\bibitem{Barenco1995} A.~Barenco, Phys. Rev. A, \textbf{52}, 3457 (1995)
\bibitem{Kitaev95} A.~Y.~Kitaev, Russ. Math. Surv., \textbf{52}, 1191, (1997)
\bibitem{Essen1955} L.~Essen, J.~V.~L. Parry, Nature \textbf{176}, 280 (1955)
\bibitem{Atom Photon interactions} C.~Cohen-Tannoudji, J.~Dupont-Roc and G.~Grynberg, {\it ``Atom-Photon Interactions''}, John Wiley and Sons inc., 1998
\bibitem{Rabbi1947} J.~E.~Nafe, E.~B.~Nelson, and I.~I.~Rabi, Phys. Rev. \textbf{71}, 914 (1947)
\bibitem{CiracZoller1995} J.~I.~Cirac and P.~Zoller, Phys. Rev. Lett. {\bf 74}, 4091 (1995)
\bibitem{SM1999} A.~S\o rensen and K.~M\o lmer, Phys. Rev. Lett. \textbf{82}, 1971 (1999)
\bibitem{Milburn2000} G.~J.~Milburn, S.~Schneider and D.~F.~V.~James, Fortschr. Phys. \textbf{48}, 801 (2000)
\bibitem{Ripoll2003} J.~J.~Garcia-Ripoll, P.~Zoller and J.~I.~Cirac, Phys. Rev. Lett., \textbf{91}, 157901 (2003)
\bibitem{Duan2006} S.-L.~Zhu, C.~Monroe and L.~M.~Duan, Phys. Rev. Lett., \textbf{97}, 050505 (2006)
\bibitem{Blatt2008} K.~Kim \emph{et al.}, Phys. Rev. A \textbf{77}, 050303 (2008)
\bibitem{Liebfried2003} D.~Liebfried \emph{et al.}, Nature, \textbf{422}, 412 (2003)
\bibitem{BenhelmNature2008} J.~Benhelm, G.~Kirchmair, C.~F.~Roos, and R.~Blatt, Nature Physics \textbf{4}, 463 (2008)
\bibitem{Wineland1978} D.~J.~Wineland, R.~E.~Drullinger and F.~L.~Walls, Phys. Rev. Lett. \textbf{40}, 1639 (1978)
\bibitem{Itano1985} W.~M.~Itano, J.~C.~Bergquist and D.~J.~Wineland, J. Opt. Soc. Am. B \textbf{2}, 1392 (1985)
\bibitem{Bollinger1985} J.~J.~Bollinger, J.~S.~Wells, D.~J.~Wineland and W.~M.~Itano, Phys. Rev. A \textbf{31}, 2711 (1985)
\bibitem{Schubert1989} M.~Schubert, I.~Siemers and R.~Blatt, J. Opt. Soc. Am. \textbf{B6}, 2159 (1989)
\bibitem{Madej1990} A.~A.~Madej and J.~D.~Sankey, Opt. Lett., \textbf{15}, 634 (1990)
\bibitem{Klein1990} H.~A.~Klein, A.~S.~Bell, G.~P.~Barwood, P.~Gill, App. Phys. B, \textbf{50}, 13 (1990)
\bibitem{Gudjons1995} T.~Gudjons, F.~Arbes, M.~Benzing, F.~Kurth and G.~Werth, Physica Scripta. \textbf{T59}, 396, (1995)
\bibitem{Tanaka1997} U.~Tanaka, H.~Imajo, K.~Hayasaka, R.~Ohmukai, M.~Watanabe, and S.~Urabe, Opt. Lett., \textbf{22}, 1353 (1997)
\bibitem{Matsubara2003} K.~Matsubara, U.~Tanaka, H.~Imajo, S.~Urabe and M.~Watanabe, App. Phys. B, \textbf{76}, 209 (2003)
\bibitem{Lucas2003} D.~M.~Lucas \emph{et al.}, Phil. Trans. R. Soc. Lond. A, \textbf{361}, 1401 (2003)
\bibitem{Wineland1998} D.~J.~Wineland \emph{et al.}, Fortschr. Phys. \textbf{46}, 363 (1998)
\bibitem{Blinov2004} B.~B.~Blinov, D.~Leibfried, C.~Monroe and D.~J.~Wineland, Quant. Inf. Proc., \textbf{3}, 45 (2004)
\bibitem{Balzer2006} C.~Balzer \emph{et al.}, Phys. Rev. A, \textbf{73}, 041407 (2006)
\bibitem{Olmschenk2007} S.~Olmschenk, K.~C.~Younge, D.~L.~Moehring, D.~N.~Matsukevich, P.~Maunz, and C.~Monroe, Phys. Rev. A \textbf{76}, 052314 (2007)
\bibitem{Benhelm2008} J.~Benhelm, G.~Kirchmair, C.~F.~Roos, and R.~Blatt, Phys. Rev. A, \textbf{77}, 062306 (2008)
\bibitem{Foot atomic physics} C.~J.~Foot, \emph{``Atomic Physics''}, Oxford University Press, 2005
\bibitem{Dehmelt1975} H.~G.~Dehmelt, Bull. Am. Phys. Soc. \textbf{20}, 60 (1975)
\bibitem{Itano1993} W.~M.~Itano \emph{et al.}, Phys. Rev. A, \textbf{47}, 3554 (1993)
\bibitem{Bergquist1987} J.~C.~Bergquist, W.~M.~Itano, and D.~J.~Wineland, Phys. Rev. A \textbf{36}, 428 (1987)
\bibitem{SchmidtKahler2003} F.~Schmidt-Kahler \emph{et al.}, J. Phys. B, \textbf{36}, 623 (2003)
\bibitem{Myerson2008} A.~Myerson \emph{et al.}, Phys. Rev. Lett. \textbf{100}, 200502 (2008)
\bibitem{Acton2006} M.~Acton, K.-A.~Brickman, P.~C.~Haljan, P.~J.~Lee, L.~Deslauriers, C.~Monroe, Quantum Inf. Comp. \textbf{6}, 465 (2006)
\bibitem{Langer2006} C.~E.~Langer, Ph. D. thesis, Department of Physics, University of Colorado, Boulder, (2006)
\bibitem{McDonnell2004} M.~J.~McDonnell \emph{et al.}, Phys. Rev. Lett., \textbf{93}, 153601 (2004)
\bibitem{Wunderlich2007} C.~Wunderlich, J. Mod. Opt., \textbf{54}, 1541 (2007)
\bibitem{Anna2011} A.~Keselamn, Y.~Glickman, N.~Akerman, S.~Kotler and R.~Ozeri New J. Phys. In Press, 	arXiv:1103.5253v1 (2011)
\bibitem{Schaetz2005} T.~Schaetz \emph{et al.}, Phys. Rev. Lett. \textbf{94}, 010501 (2005)
\bibitem{Hume2007} D.~B.~Hume, T.~Rosenband and D.~J.~Wineland, Phys. Rev. Lett. \textbf{99}, 120502 (2007).
\bibitem{Wineland LesHouches2003} D.~J.~Wineland, \emph{``Quantum entanglement and Information Processing''}, pp. 261-294. Edited by D. Est\`{e}ve, J. M. Raimond, and J. Dalibard. Proceeding of the Summer School in Les Houches, Session LXXIX, 2003. Elsevier, 2004
\bibitem{Wunderlich2001} F.~Mintert and C.~Wunderlich, Phys. Rev. Lett. \textbf{87}, 257904 (2001)
\bibitem{Chiaverini2008} J.~Chiaverini and W.~E.~Lybarger, Jr., Phys. Rev. A \textbf{77}, 022324 (2008)
\bibitem{Ospelkaus2008} C.~Ospelkaus, C.~E.~Langer, J.~M.~Amini, K.~R.~Brown, D.~Leibfried, and D.~J.~Wineland, Phys. Rev. Lett. \textbf{101}, 090502 (2008)
\bibitem{Wunderlich2009} M.~Johanning, A.~Braun, N.~Timoney, V.~Elman, W.~Neuhauser, and C.~Wunderlich, Phys. Rev. Lett. \textbf{102}, 073004 (2009)
\bibitem{Ospelkaus2011} C.~Ospelkaus, U.~Warring, Y.~Colombe, K.~R.~Brown, J.~M.~Amini, D.~Leibfried, and D.~J.~Wineland, arXiv:1104.3573v2 (2011)
\bibitem{Ozeri2007} R.~Ozeri \emph{et. al.}, Phys. Rev. A \textbf{75}, 042329 (2007)
\bibitem{Ozeri2005} R.~Ozeri \emph{et. al.}, Phys. Rev. Lett. \textbf{95}, 030403 (2005)
\bibitem{Uys2010} H.~Uys \emph{et. al.}, Phys. Rev. Lett. \textbf{105}, 200401 (2005)
\bibitem{Berry1984} M.~V.~Berry, Proc. Roy. Soc. Lon., \textbf{392}, 45, (1984)
\bibitem{AharonovAnandan1987} Y.~Aharonov and J.~Anandan, Phys. Rev. Lett., \textbf{58}, 1593 (1987)
\bibitem{CiracZoller2000} J.~I.~Cirac and P.~Zoller, Nature {\bf 404}, 579 (2000)
\bibitem{Zhu2003} S.-L.~Zhu and Z.~D.~Wang, Phys. Rev. Lett. \textbf{91}, 187902 (2003)
\bibitem{Lee2005} P. J. Lee, \emph{et. al.}, Journal of Optics B, \textbf{7}, S371 (2005)
\bibitem{DFV James} D.~F.~V.~James, Applied Physics B \textbf{66}, 181 (1998)
\bibitem{Gate squeezing Oxford} M.~J.~McDonnell \emph{et. al.}, Phys. Rev. Lett. \textbf{98}, 063603 (2007)
\bibitem{Haljan2005} P.~C.~Haljan, \emph{et. al.}, Phys. Rev. A \textbf{72}, 062316 (2005)
\bibitem{Aharonov1996} D.~Aharonov and M.~Ben-Or, arXiv:quant-ph/9611025v2 (1996)
\bibitem{Preskill1998} J.~Preskill, Proc. R. Soc. Lond. A  \textbf{454}, 385 (1998)
\bibitem{Knill2005} E.~Knill, Nature \textbf{434}, 39 (2005)
\bibitem{Vandevender2010} A.~VanDevender, Y. Colombe, J. Amini, D. Leibfried, and D. J. Wineland, Phys. Rev. Lett. \textbf{105}, 023001 (2010)
\bibitem{Knill2008} E.~Knill, \emph{et. al.}, Phys. Rev. A \textbf{77}, 012307 (2008)
\bibitem{Brown2011} K.~R.~Brown, \emph{et. al.}, arXiv:1104.2552v1 (2011)
\end{thebibliography}
\end{document}